\documentclass[aps,prx,reprint,amsmath,amssymb,superscriptaddress,floatfix,longbibliography]{revtex4-2}

\usepackage{graphicx}
\usepackage{bm}
\usepackage{color}
\usepackage[normalem]{ulem}
\usepackage{txfonts}
\usepackage[colorlinks=true, urlcolor=blue, linkcolor=blue, citecolor=blue]{hyperref}
\usepackage{physics}

\newcommand{\longbar}[1]{\overline{#1}}

\begin{document}

\title{Frustration from Localized Zhang-Rice States: \\
A Unified Theory of Doping-Driven Magnetic Transitions in Cuprates}

\author{Xiaodong Wang}
\thanks{These authors contributed equally.}
\affiliation{Beijing National Laboratory for Condensed Matter Physics and \\ Institute of Physics,
	Chinese Academy of Sciences, Beijing 100190, China}
\affiliation{School of Physical Sciences, University of Chinese Academy of Sciences, Beijing 100190, China}

\author{Ping Xu}
\thanks{These authors contributed equally.}
\affiliation{Key Laboratory of Artificial Structures and Quantum Control (Ministry of Education), School of Physics and Astronomy, Shanghai Jiao Tong University, Shanghai 200240, China}

\author{Jiong Mei}
\affiliation{Beijing National Laboratory for Condensed Matter Physics and \\ Institute of Physics,
	Chinese Academy of Sciences, Beijing 100190, China}
\affiliation{School of Physical Sciences, University of Chinese Academy of Sciences, Beijing 100190, China}

\author{Shao-Hang Shi}
\affiliation{Beijing National Laboratory for Condensed Matter Physics and \\ Institute of Physics,
	Chinese Academy of Sciences, Beijing 100190, China}
\affiliation{School of Physical Sciences, University of Chinese Academy of Sciences, Beijing 100190, China}

\author{Zi-Xiang Li}
\email{zixiangli@iphy.ac.cn}
\affiliation{Beijing National Laboratory for Condensed Matter Physics and \\ Institute of Physics,
	Chinese Academy of Sciences, Beijing 100190, China}
\affiliation{School of Physical Sciences, University of Chinese Academy of Sciences, Beijing 100190, China}

\author{Mingpu Qin}
\email{qinmingpu@sjtu.edu.cn}
\affiliation{Key Laboratory of Artificial Structures and Quantum Control (Ministry of Education), School of Physics and Astronomy, Shanghai Jiao Tong University, Shanghai 200240, China}
\affiliation{Hefei National Laboratory, Hefei 230088, China}

\author{Kun Jiang}
\email{jiangkun@iphy.ac.cn}
\affiliation{Beijing National Laboratory for Condensed Matter Physics and \\ Institute of Physics,
	Chinese Academy of Sciences, Beijing 100190, China}
\affiliation{School of Physical Sciences, University of Chinese Academy of Sciences, Beijing 100190, China}

\author{Hui-Ke Jin}
\email{jinhk@shanghaitech.edu.cn}
\affiliation{School of Physical Science and Technology, ShanghaiTech University, Shanghai 201210, China}

\date{\today}

\begin{abstract}
The microscopic mechanism by which doped holes disrupt the antiferromagnetic (AFM) order is one of the fundamental questions in cuprates.
In this work, we propose a unified microscopic theory in which doped holes form spatially localized Zhang-Rice singlets which actively mediate emergent spin exchange. Rather than acting as simple non-magnetic vacancies, these localized states introduce emergent next-nearest $J_2$ and third-nearest $J_3$ neighbor superexchanges. This dopant-induced exchange pathway generates significant magnetic frustration, naturally explaining the rapid collapse of the N\'eel AFM order and the emergence of a spin-glass phase on the hole-doped side. In contrast, electron doping primarily leads to spin dilution, allowing AFM order to persist to substantially higher doping levels, consistent with the experimentally observed electron-hole asymmetry.
Our findings provide a unified microscopic framework for understanding doping-driven magnetic transitions in lightly doped cuprates.

\end{abstract}

\maketitle

\section{Introduction}


Carrier doping transforms the antiferromagnetic (AFM) charge-transfer insulating parent state of cuprates into a rich sequence of correlated phases, eventually giving rise to superconductivity~\cite{Lee2006,Keimer2015,Norman2003,Bednorz1986,Armitage2010}. Understanding how doped carriers reshape this strongly correlated background is therefore a central challenge both experimentally and theoretically. A particularly striking manifestation of this doping process is the pronounced electron-hole asymmetry in magnetism: while long-range N\'eel order persists up to an electron doping of $\delta \approx 0.15$~\cite{Luke1990,Keimer1992,Matsuda1992,Armitage2010}, it is rapidly destroyed at only $\delta_c \approx 0.02$--$0.05$ upon hole doping~\cite{Chou1993,Niedermayer1998,Kastner1998}. Uncovering the microscopic origin of this fundamental asymmetry remains crucial for a unified understanding of the cuprate phase diagram on both the electron- and hole-doped sides.

\begin{figure}
    \centering    \includegraphics[width=1.0\linewidth]{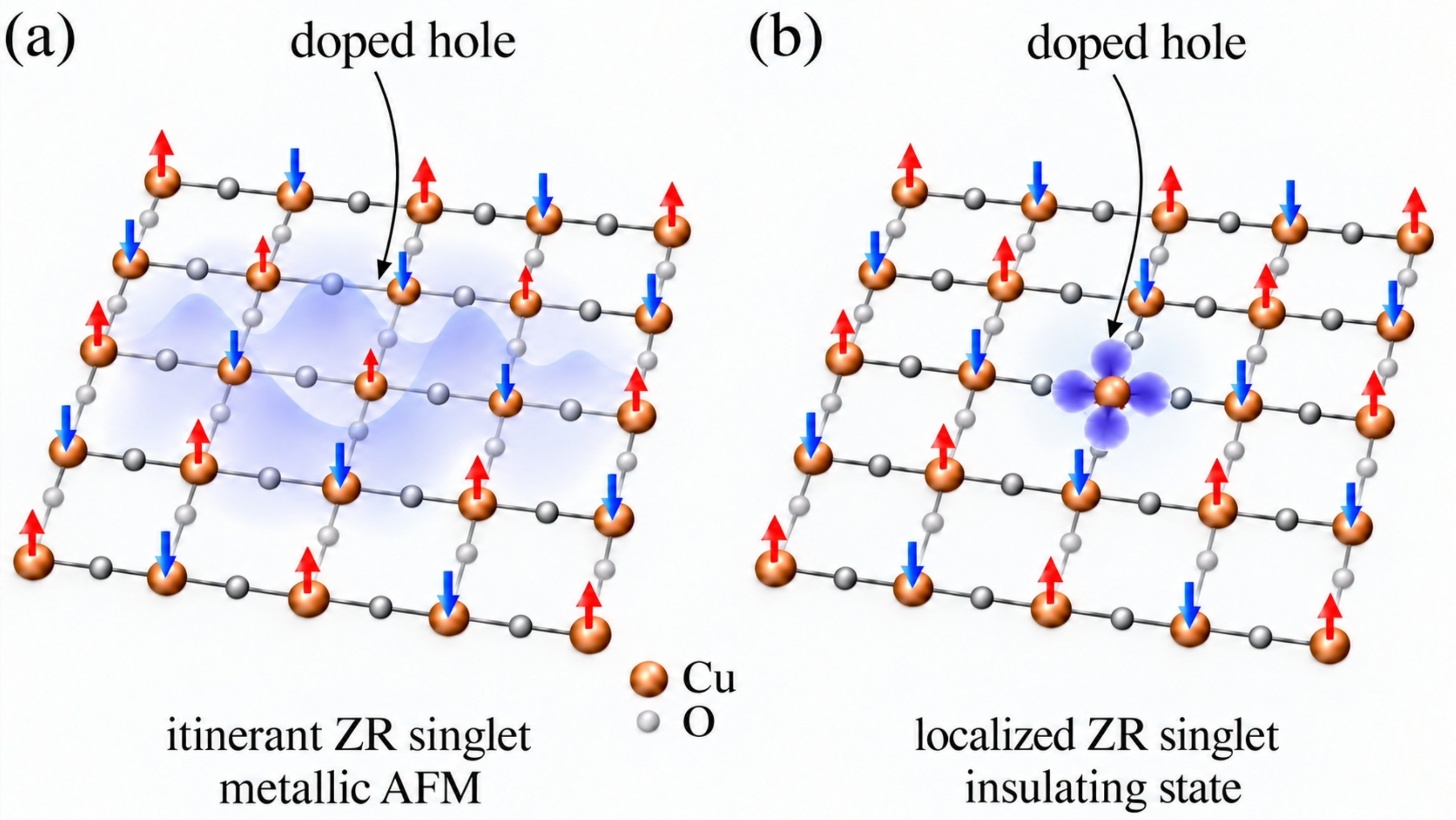}
    \caption{Schematic contrast between clean and disorder-localized hole doping in a CuO\(_2\) antiferromagnet. (a) In the clean limit, the doped hole forms an itinerant Zhang-Rice (ZR) singlet that delocalizes over multiple plaquettes and supports a metallic AFM state. (b) With disorder such as dopant potential, the doped hole is localized into a ZR singlet, producing an insulating state.}
    \label{fig:zrsketch}
\end{figure}

In the clean limit, Hubbard and $t$–$J$ model frameworks predict a metallic AFM state, in which doped holes form itinerant Zhang-Rice singlets and a Fermi surface coexists with long-range AFM order~\cite{Zhang1988PRB,Emery1987,Shraiman1988,Kane1989,SchmittRink1988,Su1989,Sachdev1989,Sheng1996,Weng1997,Weng2007}; as illustrated in Fig.~\ref{fig:zrsketch}(a). This scenario has been strikingly supported by recent ARPES and quantum oscillation experiments on the disorder-shielded inner planes of multilayer cuprates, which observe well-defined small Fermi pockets within a robust AFM background~\cite{Kunisada2020,Chen2026,Jeong2026}. In stark contrast, in most chemically doped cuprates, this clean-limit metallic AFM state is absent; instead, light hole doping yields an insulating AFM parent state that rapidly collapses into an insulating spin-glass regime.
This stark discrepancy directly underscores the pivotal role of dopant-induced spatial disorder~\cite{Aharony1988,Emery1988,Keimer1992,EMERY1993597,pan2001,Wang2002,Shraiman1989,Alloul2009}. Indeed, local probes such as scanning tunneling microscopy (STM) reveal that out-of-plane dopant potentials locally trap charge carriers, forming spatially localized ZR singlets rather than fully itinerant carriers~\cite{Ye2023,Li2024,Mei_2025,zhao_2025,xia2025,Wang_2025,Chen2025arXivA}; see the schematic illustration in Fig.~\ref{fig:zrsketch}(b).
This localization poses a fundamental paradox: if kinetic carrier motion is quenched into static localized ZR states by disorder, how do these static states so aggressively destroy long-range AFM order and drive the system into an insulating spin glass?

Resolving this paradox requires understanding how localized ZR singlets modify the magnetic interactions among the surrounding Cu spins.
Early theories proposed that bare oxygen holes might introduce magnetic frustration effects via local ferromagnetic bonds~\cite{Aharony1988,Emery1988} or impurities~\cite{Liu2009,Liu2013,Harter2007,Andersen2007}.
Here we reveal a fundamentally different mechanism. 
We demonstrate that the fully formed, localized ZR singlets act as active intermediate states for virtual hopping. This dopant-mediated channel generates emergent longer-range superexchange interactions ($J_2$ and $J_3$) between  copper spins. 


Employing large-scale Monte Carlo simulations and density matrix renormalization group (DMRG) calculations~\cite{White1992,White1993}, we demonstrate that this dopant-induced frustration strongly disrupts the N\'eel background, naturally explaining the rapid collapse of the AFM order and the subsequent emergence of a robust spin-glass phase upon hole doping~\cite{Chou1993,Niedermayer1998,Gooding1997,Matsuda2000PRB}. When contrasted with the simple spin-dilution mechanism in electron-doped cuprates~\cite{Keimer1992,Matsuda1992,Mei_2025}, our results provide a unified perspective on the role of dopants. In particular, we identify the dopant potential as the microscopic origin of the pronounced magnetic electron-hole asymmetry---a long-standing puzzle highlighted in early theories~\cite{Schmalian1993,Tohyama1994,Gooding1994,Gooding1994_2} and reviews~\cite{Armitage2010}---as recently suggested in our previous work~\cite{Mei_2025}.

The rest of this paper is organized as follows. 
In Sec.~\ref{sec:3band}, we introduce the microscopic framework starting from the three-band Emery model with a local dopant potential. Sec.~\ref{sec:j1j2} details how localized ZR singlets mediate emergent longer-range $J_2$ and $J_3$ superexchanges, based on which the effective spin model is constructed in Sec.~\ref{sec:spin_model}. 
We present the zero-temperature quantum phase diagram obtained via large-scale VMC and DMRG calculations in Sec.~\ref{sec:ground_state}. Specifically, Sec.~\ref{sec:Mz_Sq} demonstrates the magnetic electron-hole asymmetry by evaluating the static spin structure factor and staggered magnetization, while Sec.~\ref{sec:domain_wall} reveals the real-space domain-wall structures driving the N\'eel order collapse. 
In Sec.~\ref{sec:finite_T}, we extend our analysis to a 3D layered geometry using classical Monte Carlo simulations. Sec.~\ref{sec:TN} evaluates the doping dependence of the N\'eel temperature $T_N$ using Binder cumulative scaling, and Sec.~\ref{sec:SG} details the emergence of a 3D spin-glass phase once the N\'eel order is destroyed. Finally, Sec.~\ref{sec:discussion} provides a summary and discusses experimental implications.

\begin{figure*}[t]
  \centering
  \includegraphics[width=0.98\textwidth]{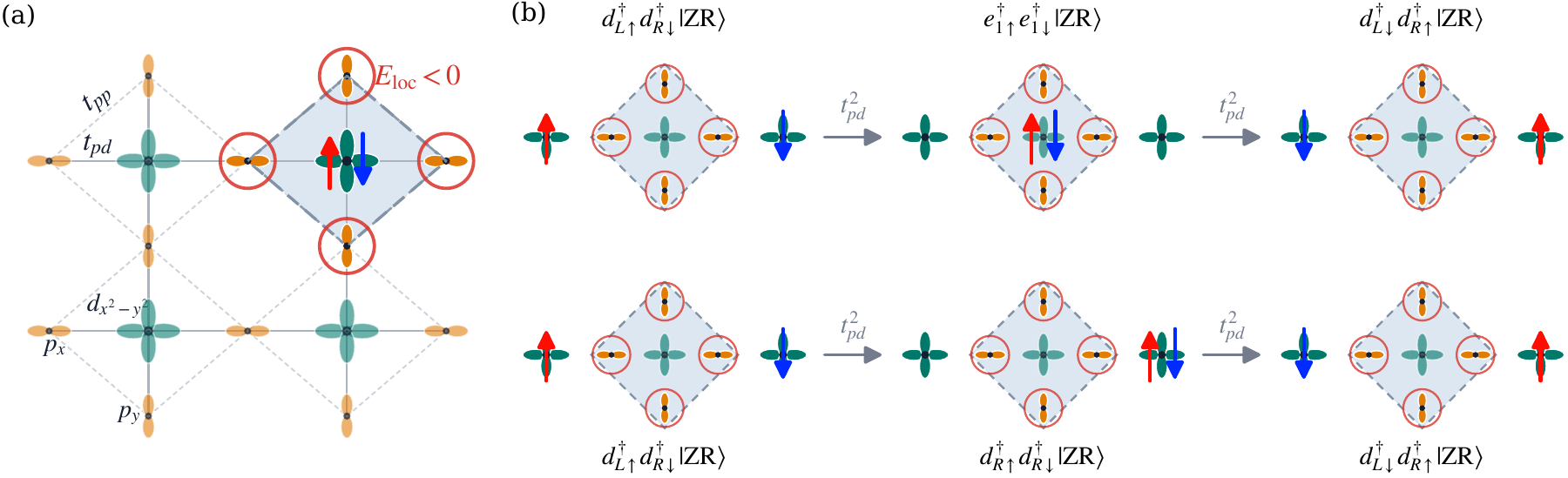}
  \caption{Localized Zhang--Rice (ZR) singlet and the associated virtual exchange processes in the three-band Emery model. (a) Cu $d_{x^2-y^2}$ and O $p_{x/y}$ orbitals in the CuO$_2$ plane, with Cu--O and O--O hopping magnitudes $t_{pd}$ and $t_{pp}$, respectively; their bond-dependent signs, fixed by the orbital phases, are not displayed. The local attractive potential $E_{\rm loc}<0$ lowers the oxygen levels on the highlighted CuO$_4$ plaquette and localizes the ZR singlet. Red and blue arrows denote spin-up and spin-down holes, respectively. (b) Two representative fourth-order paths that exchange the spins on the outer Cu orbitals $d_L$ and $d_R$ while the localized $|{\rm ZR}\rangle$ state remains the reference state. The upper path passes through the oxygen-doublon state $e_{1\uparrow}^{\dagger}e_{1\downarrow}^{\dagger}|{\rm ZR}\rangle$, whereas the lower path passes through the Cu-doublon state $d_{R\uparrow}^{\dagger}d_{R\downarrow}^{\dagger}|{\rm ZR}\rangle$. Each gray arrow groups two Cu--O hopping events, giving the factor $t_{pd}^{2}/2$ after projection onto the $e_1$ channel.}
  \label{fig:ZR}
\end{figure*}

\section{Microscopic Model}\label{sec:microscopic_model}
\subsection{Three-Band Emery Model and Localized Dopant State}\label{sec:3band}

Microscopically, the essential physics of cuprates originates from their charge-transfer nature, which can be captured by the three-band Emery model in the hole representation~\cite{Emery1987,Zhang1988PRB}. As illustrated in Fig.~\ref{fig:ZR}(a), we choose the Cu $d_{x^2-y^2}$ level as the zero of energy and denote by $d_{i\sigma}^{\dagger}$ and $p_{l\sigma}^{\dagger}$ the creation operators for a hole with spin
$\sigma$ on Cu site $i$ and O site $l$, respectively. The corresponding number operators are $\hat n_{i\sigma}^{d}=d_{i\sigma}^{\dagger}d_{i\sigma}$ and $\hat n_{l\sigma}^{p}=p_{l\sigma}^{\dagger}p_{l\sigma}$. In this convention, the microscopic Hamiltonian reads
\begin{equation}
\begin{split}
H_{\rm 3band}={}&\xi_p\sum_{l,\sigma}\hat n_{l\sigma}^{p}
+U_d\sum_i\hat n_{i\uparrow}^{d}\hat n_{i\downarrow}^{d}
+U_p\sum_l\hat n_{l\uparrow}^{p}\hat n_{l\downarrow}^{p}\\
+t_{pd}&\sum_{\langle il\rangle,\sigma}
\left(d_{i\sigma}^{\dagger}p_{l\sigma}+\mathrm{h.c.}\right)
+t_{pp}\sum_{\langle ll'\rangle,\sigma}
\left(p_{l\sigma}^{\dagger}p_{l'\sigma}+\mathrm{h.c.}\right).
\end{split}
\label{eq:three_band_main}
\end{equation}
Here, $\xi_p>0$ is the charge-transfer energy, and $U_d$ asl well as $U_p$ are the on-site Coulomb repulsions on the Cu and O orbitals, respectively. The parameters $t_{pd}$ and $t_{pp}$ denote the hopping matrix for the nearest-neighbor Cu--O and O--O processes. 
The bond-dependent signs of the hopping matrix elements, arising from the orbital phases, are left implicit in Eq.~\eqref{eq:three_band_main} and Fig.~\ref{fig:ZR}(a). These signs are fixed once an orbital phase convention is chosen; for instance, different but physically equivalent conventions are used in Refs.~\cite{White2015,Cui2020}. When the hopping matrix elements and orbital basis are transformed consistently, the resulting exchange couplings are independent of the gauge choice and the explicit forms of bond sign.

In the following, we use $t_{pd}$ and $t_{pp}$ to denote the positive hopping magnitudes. For the cuprates, the parameters in Eq.~\eqref{eq:three_band_main} satisfy $U_d\gg \xi_p > t_{pd}\gg t_{pp}$, placing the system in the charge-transfer regime~\cite{Zaanen1985}. The representative parameter values are chosen as $t_{pd}\simeq 1.1$~eV, $t_{pp}\simeq 0.55$~eV, $\xi_p\simeq 3.3$~eV, and $U_d\simeq 8.8$~eV. For simplicity, we use $U_p=0$ below, which does not qualitatively affect the results.

In the undoped limit, the strong Coulomb repulsion localizes one hole per unit cell predominantly on the Cu site. Virtual charge hopping through the intervening O orbitals then generates a nearest-neighbor (NN) AFM superexchange between the localized spin-$1/2$ Cu moments, stabilizing the N\'eel-ordered ground state~\cite{Zhang1988PRB,Anderson1959}. To leading order, this exchange is captured by the effective Heisenberg Hamiltonian $H_{\rm Heisenberg} = J_1 \sum_{\langle ij\rangle} \mathbf{S}_i \cdot \mathbf{S}_j,$
where $\mathbf{S}_i$ denotes the vector of three spin-$1/2$ operators at site $i$. The effective NN exchange coupling is given by $J_1 \approx \frac{4t_{pd}^4}{\xi_p^2}\left(\frac{1}{U_d}+\frac{1}{\xi_p}\right)$~\cite{Zaanen1985}; its perturbative derivation is provided in Appendix~\ref{app:downfolding}.

Upon hole doping, the additional holes  reside predominantly on the O orbitals and bind with the Cu moments to form ZR singlets~\cite{Zhang1988PRB}. In real materials, however, chemical substitution or oxygen nonstoichiometry also generates a spatially inhomogeneous electrostatic potential. In the hole representation, a locally attractive component of this potential lowers the ligand-hole energy and can pin a ZR singlet to a particular CuO$_4$ plaquette \cite{Li2024}. As illustrated by the highlighted plaquette in Fig.~\ref{fig:ZR}(a), the doped hole is distributed over the four O orbitals surrounding the central Cu site.
We denote the plauqette formed by those four O sites by $\mathcal P_0$ and retain the leading $D_4$-symmetric component of the dopant potential. The local Hamiltonian to mimic the effect of the dopant is
\begin{equation}
H_{\rm loc}=E_{\rm loc}\sum_{l\in\mathcal P_0,\sigma}\hat n_{l\sigma}^{p},
\qquad E_{\rm loc}<0,
\label{eq:local_dopant_potential}
\end{equation}
where $E_{\rm loc}$ is the energy shift per O orbital within $\mathcal P_0$. Eq.~\eqref{eq:local_dopant_potential} lowers the local charge-transfer energy on the selected plaquette and thereby favors spatial localization of the ZR singlet. For the quantitative estimates below, we adopt the phenomenological representative value $E_{\rm loc}=-2$~eV. The effective local charge-transfer energy, e.g., $\xi_e = \xi_p + E_{\rm loc} \approx 1.3$~eV, is reduced by a factor of $2$--$3$ compared to the bare charge-transfer gap ($\xi_p \simeq 3.3$~eV). This dopant-induced reduction naturally explains the low-energy in-gap states observed in recent STM experiments~\cite{Ye2023,Li2024,Wang_2025}. 

The reduction of the local charge-transfer energy binds the doped hole to $\mathcal P_0$, stabilizing a spatially localized ZR singlet rather than an itinerant state extending throughout the CuO$_2$ plane. To capture the physics of this localized state, we exploit the $D_4$ point-group symmetry of the highlighted CuO$_4$ plaquette in Fig.~\ref{fig:ZR}(a) to construct a natural basis for a local-cluster description. Denoting the hole creation operators on the four surrounding oxygen sites as $p_{\pm\hat{\mathbf x},\sigma}^\dagger$ and $p_{\pm\hat{\mathbf y},\sigma}^\dagger$, we define the symmetry-adapted orthonormal combinations:
\begin{equation}
\begin{aligned}
&a^\dagger_\sigma=\frac{1}{2}\left(p_{\hat{\mathbf x},\sigma}^\dagger+p_{\hat{\mathbf y},\sigma}^\dagger-p_{-\hat{\mathbf x},\sigma}^\dagger-p_{-\hat{\mathbf y},\sigma}^\dagger\right),\\
&b^\dagger_\sigma=\frac{1}{2}\left(p_{\hat{\mathbf x},\sigma}^\dagger-p_{\hat{\mathbf y},\sigma}^\dagger-p_{-\hat{\mathbf x},\sigma}^\dagger+p_{-\hat{\mathbf y},\sigma}^\dagger\right),\\
&e_{1\sigma}^\dagger=\frac{1}{\sqrt{2}}\left(p_{\hat{\mathbf x},\sigma}^\dagger+p_{-\hat{\mathbf x},\sigma}^\dagger\right),\quad
e_{2\sigma}^\dagger=\frac{1}{\sqrt{2}}\left(p_{\hat{\mathbf y},\sigma}^\dagger+p_{-\hat{\mathbf y},\sigma}^\dagger\right).
\label{eq:ligand_orbitals}
\end{aligned}
\end{equation}
These ligand orbitals diagonalize the intra-plaquette O--O hopping. When dressed by the local dopant potential in Eq.~\eqref{eq:local_dopant_potential}, their effective energies become:
\begin{equation}
\begin{aligned}
\xi_b&=\xi_p-2t_{pp}+E_{\rm loc},\qquad
\xi_a=\xi_p+2t_{pp}+E_{\rm loc},\\
\xi_{e_1}&=\xi_{e_2}=\xi_p+E_{\rm loc}.
\end{aligned}
\label{eq:ligand_energies}
\end{equation}
Crucially, among the symmetry-adapted O orbitals in Eq.~\eqref{eq:ligand_orbitals}, only the $b^\dagger$ orbital transforms with the same spatial symmetry as the central Cu $d_{x^2-y^2}$ orbital. Consequently, the intra-plaquette Cu--O hopping involving the central Cu orbital reduces to $2t_{pd}\sum_\sigma(d_\sigma^\dagger b_\sigma+\mathrm{h.c.})$. Within the ideal $D_4$-symmetric plaquette, the corresponding matrix elements between the central Cu orbital and the $a$, $e_1$, or $e_2$ orbitals vanish by symmetry. Driven by this strong hybridization and the fact that the lowest oxygen-orbital energy $\xi_b$ is further deepened by the attractive potential $E_{\rm loc}<0$, the doped oxygen hole preferentially occupies the $b$ orbital.

The local two-hole sector consists of the intrinsic Cu hole and the doped hole in the $b$ orbital. Crucially, these two holes strongly bind into a local singlet, effectively quenching the central Cu moment and removing it from the low-energy spin sector. 
For $U_d\gg \xi_b,t_{pd}$, the high-energy Cu-doublon configuration can be integrated out. The low-energy physics is then governed by two singlet configurations: the Cu--O singlet $|S_{db}\rangle=\frac{1}{\sqrt{2}}(b_\downarrow^\dagger d_\uparrow^\dagger-b_\uparrow^\dagger d_\downarrow^\dagger)|0\rangle$ and the $b$-orbital doublon $|S_{bb}\rangle=b_\downarrow^\dagger b_\uparrow^\dagger|0\rangle$. In the basis $(|S_{db}\rangle,|S_{bb}\rangle)^T$, the effective Hamiltonian is
\begin{equation}
H_{\rm ZR}=
\begin{pmatrix}
\xi_b-8t_{pd}^2/U_d & 2\sqrt{2}t_{pd}\\
2\sqrt{2}t_{pd} & 2\xi_b
\end{pmatrix}.
\label{eq:zr_local_hamiltonian}
\end{equation}
The degree of mixing is determined by the diagonal mismatch $\Delta_{\rm ZR}=\xi_b+8t_{pd}^2/U_d$ relative to the hybridization $2\sqrt{2}t_{pd}$. For typical cuprate parameters and a dopant potential $E_{\rm loc}=-2.0$~eV, the mismatch ($\Delta_{\rm ZR} \approx 1.3$~eV) is overwhelmed by the effective hybridization ($4\sqrt{2}t_{pd} \approx 6.2$~eV). To capture the essential physics in the subsequent analysis, we assume the strong-mixing limit $\Delta_{\rm ZR} \ll 2\sqrt{2}t_{pd}$, where the low-energy eigenstate of Eq.~\eqref{eq:zr_local_hamiltonian} simplifies to
\begin{equation}
|{\rm ZR}\rangle\simeq
\frac{1}{\sqrt{2}}\left[|S_{db}\rangle-|S_{bb}\rangle\right]
\label{eq:localized_zr_state}
\end{equation}
As a superposition of a Cu--O singlet and a $b$-orbital doublon, $|{\rm ZR}\rangle$ will serve below as the reference state for virtual charge fluctuations through the remaining oxygen orbitals.

\subsection{Dopant-Mediated $J_2$ and $J_3$ Exchanges}\label{sec:j1j2}

While the localized ZR singlet removes the central Cu moment from the low-energy spin sector, it fundamentally differs from an inert non-magnetic vacancy. Because the lightly doped system remains insulating, charge transfer into and out of the ZR plaquette occurs only via virtual processes. The relevant low-energy manifold thus consists of a fixed $|{\rm ZR}\rangle$ state on $\mathcal P_0$ alongside one spin-$1/2$ hole on each surrounding Cu site. Integrating out the high-energy charge fluctuations around this manifold generates emergent $J_2$ and $J_3$ superexchange interactions between Cu spins that are spatially separated beyond the nearest-neighbor bonds of the undoped lattice.

To elucidate how the emergent $J_2$ and $J_3$ couplings arise, we first examine a representative $J_3$ pathway. Consider two Cu orbitals, $d_L$ and $d_R$, located on opposite sides of the ZR plaquette along the $x$ direction, as illustrated in Fig.~\ref{fig:ZR}(b). These orbitals form a third-nearest-neighbor (3rd-NN) pair on the Cu lattice. 
The outer Cu orbitals $d_L$ and $d_R$ couple to individual O orbitals of the plaquette and therefore project onto several symmetry-adapted orbital channels. Reexpressing the Cu--O hopping in terms of the oxygen orbitals defined in Eq.~\eqref{eq:ligand_orbitals} yields $H_{\rm hyb}^{(3)}=t_{pd}\sum_\sigma [ d_{L\sigma}^{\dagger} ( -e_{1\sigma}/\sqrt{2} + a_\sigma/2 + b_\sigma/2 ) + d_{R\sigma}^{\dagger} ( e_{1\sigma}/\sqrt{2} + a_\sigma/2 + b_\sigma/2 ) ] + \mathrm{h.c.}$ Although the $b$ orbital is the lowest in energy, it already carries most of the oxygen hole weight in the $|{\rm ZR}\rangle$ state. Additional virtual hopping through this channel is therefore strongly restricted by Pauli blocking and the internal excitation gap of the ZR singlet. Consequently, the dominant superexchange instead proceeds through the initially empty $e_1$ and $e_2$ orbitals. As indicated by Eq.~\eqref{eq:ligand_energies}, the effective energy of these channels $\xi_e$ is comparable to the bare charge-transfer gap $\xi_p$ and the Hubbard repulsion $U_d$. Because they contribute energy denominators of order $1/\xi_e$ to the fourth-order perturbation, these intermediate states are energetically highly accessible and must be explicitly included in the exchange mechanism.

The mechanism of emergent longer-range exchanges is particularly transparent in the $e_1$ channel. Instead of localized on a single atomic site, the $e_1$ and $a$ states are extended molecular orbitals defined over the entire CuO$_4$ plaquette. Consequently, they can efficiently bridge distant Cu spins. This spatial extension allows them to mediate the 2nd-NN $J_2$ and 3rd-NN $J_3$ exchanges, within the exact same number of virtual hopping steps (fourth-order) as the conventional 1st NN exchange. 

For instance, Fig.~\ref{fig:ZR}(b) illustrates two representative fourth-order sequences starting from an antiparallel spin pair on $d_L$ and $d_R$. Along the upper path, the two outer holes successively hop into the extended $e_1$ orbital, forming an oxygen-doublon intermediate state $e_{1\uparrow}^{\dagger}e_{1\downarrow}^{\dagger}|{\rm ZR}\rangle$. The subsequent two hops return the holes to the opposite Cu orbitals, thereby exchanging their spins. Alternatively, along the lower path, a hole first enters $e_1$ and then hops onto the already occupied $d_R$ site, creating a Cu-doublon state $d_{R\uparrow}^{\dagger}d_{R\downarrow}^{\dagger}|{\rm ZR}\rangle$. The remaining hops transfer the original $d_R$ hole with opposite spin through $e_1$ to $d_L$. Integrating out these intermediate charge sectors yields the dominant $e_1$-channel contribution to the antiferromagnetic exchange:
\begin{equation}
J_{3e}=\frac{4t_{pd}^4}{\xi_e^2}
\left(\frac{1}{U_d}+\frac{1}{\xi_e}\right)
\simeq \frac{t_{pd}^4}{\xi_e^3},
\qquad \xi_e=\xi_p+E_{\rm loc},
\label{eq:dominant-j3-channel}
\end{equation}
where the last form applies for $U_d\gg\xi_e$. Eq.~\eqref{eq:dominant-j3-channel} reveals a striking departure from the undoped case. While the conventional $J_1$ exchange is suppressed by the large bare charge-transfer gap $\xi_p$, the dopant-mediated $J_3$ is governed by the locally reduced energy $\xi_e$. The attractive dopant potential therefore drastically amplifies the oxygen-doublon pathway, allowing the emergent $J_3$ to become comparable to, or even exceed, the nearest-neighbor $J_1$. where the last form applies for $U_d\gg\xi_e$. Equation~\eqref{eq:dominant-j3-channel} reveals a striking departure from the undoped case. The total third-nearest-neighbor exchange is given by $J_3 = J_{3e} + J_{3u}$, where $J_{3u}$ encapsulates the additional contributions from the $a$ and $b$ channels as well as their interference terms; the complete fourth-order derivation is detailed in Appendix~\ref{app:downfolding}.

\begin{figure*}[t]
  \centering
  \includegraphics[width=0.92\textwidth]{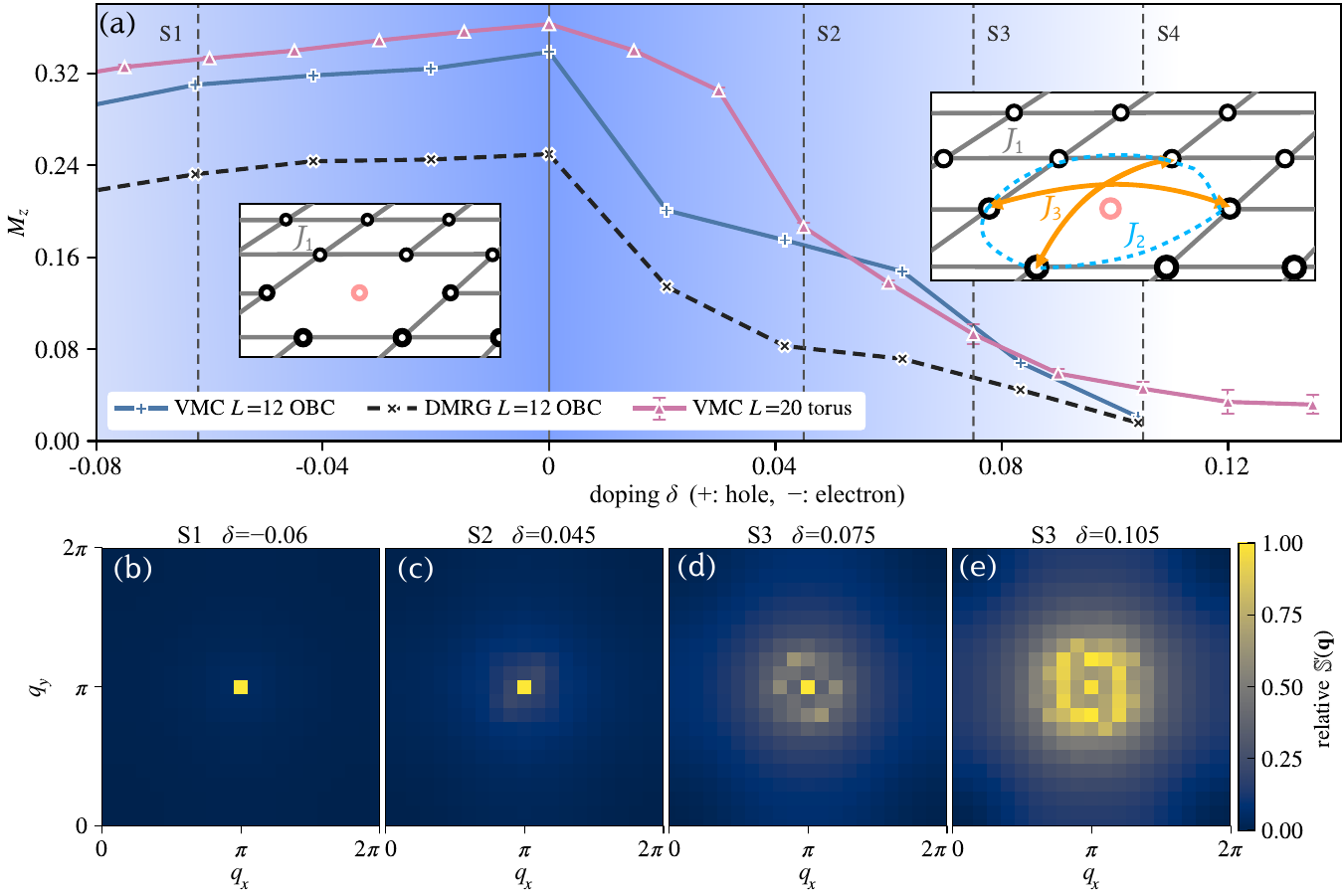}
\caption{Asymmetric evolution of the AFM order upon electron and hole doping. (a) The staggered magnetization $M_z$ in Eq.~\eqref{eq:Mz}  as a function of doping concentration. Negative and positive values on the x axis represent electron and hole doping, respectively. The error bars are comparable to the marker size. Pink triangles denote the large-scale VMC results on $L=20$ torus geometries, averaged over random configurations of dopants. Blue crosses and black dashed lines display the VMC and DMRG benchmarks on smaller $L=12$ lattices with open boundary conditions (OBC). {\bf Insets}: Schematic representations of the effective spin models for electron (left) and hole (right) doping. The dopant is indicated by a transparent red circle. In both cases, the 1st nearest-neighbor (NN) $J_1$ bonds (gray lines) connected to the dopant site are removed. For hole doping (right), frustrating 2nd NN $J_2$ (blue dashed lines) and 3rd NN $J_3$ (orange arrows) exchange pathways emerge around the dopant, which are absent in the electron-doped case. (b)-(e) Static spin structure factor $\mathbb{S}(\mathbf{q})$ obtained by VMC  for four representative doping levels marked by vertical dashed lines in (a): (b) S1 ($\delta=-0.06$), (c) S2 ($\delta=0.045$), (d) S3 ($\delta=0.075$), and (e) S4 ($\delta=0.105$). 
Here we use $J_1=1$, $J_2=0.5$, and $J_3=1.25$. }
\label{fig:fig1}
\end{figure*}

An analogous downfolding generates the 2nd-NN exchange $J_2$ between two Cu sites located on adjacent sides of the ZR plaquette, such as $d_L$ and $d_U$. The orbital geometry, however, differs fundamentally from the collinear $d_L$--$d_R$ configuration relevant to $J_3$: $d_L$ couples to the $e_1$ orbital, whereas $d_U$ couples to the orthogonal $e_2$ orbital. Because $e_1$ and $e_2$ do not mix in the $D_4$-symmetric plaquette, the two Cu sites cannot exchange through a common $e$ orbital. Consequently, the dominant $e$-orbital process is absent for $J_2$. The $J_2$ exchange is instead generated by the weaker $a$- and $b$-orbital channels, naturally yielding $J_2<J_3$. The complete expressions of $J_2$ are presented in Appendix~\ref{app:downfolding}.

\begin{table}[!t]
\centering
\caption{Comparison of the effective superexchange couplings obtained from  ED and fourth-order perturbation theory. The representative cuprate parameters are $t_{pd}=1.1$~eV, $t_{pp}=0.55$~eV, $\xi_p=3.3$~eV, $U_p=0$, $U_d=8.8$~eV, and $E_{\rm loc}=-2.0$~eV.}
\label{tab:benchmark}
\renewcommand{\arraystretch}{1.3}
\setlength{\tabcolsep}{12pt}
\begin{tabular}{c c c}
\hline\hline
Coupling & ED (eV) & Perturbative (eV) \\
\hline
$J_1$ & 0.149 & 0.224 \\
$J_2$ & 0.0239 & 0.0334 \\
$J_3$ & 0.185 & 0.384 \\
\hline
$J_1:J_2:J_3$ & 1 : 0.16 : 1.24 & 1 : 0.15 : 1.71 \\
\hline\hline
\end{tabular}
\end{table}

To quantitatively validate this emergent exchange mechanism and the resulting hierarchy, we benchmark the perturbative analytical expressions against exact diagonalization (ED) on representative 7-site clusters shown in Fig.~\ref{fig:ZR} (b). Using the typical cuprate parameters introduced above, the resulting effective couplings are summarized in Table~\ref{tab:benchmark}. The ED results confirm that the dopant-mediated 3rd-NN exchange $J_3$ is remarkably strong, slightly exceeding the bare nearest-neighbor exchange $J_1$. In contrast, the geometrically suppressed 2nd-NN exchange $J_2$ remains small. Both the absolute magnitudes and the robust hierarchy $J_3 \gtrsim J_1 > J_2$ are well captured by the fourth-order perturbation theory. This quantitative agreement firmly establishes that a localized ZR center actively generates substantial frustrating exchange across the dopant, rather than merely acting as a bond-cutting vacancy. 
Finally, we note that effective $J_2$ and $J_3$ couplings can also arise from a different mechanism, where Zn--O hybridization creates a low-lying impurity state below the Hubbard scale and thereby enhances further-neighbor exchange processes~\cite{Liu2009,Liu2013}.


\subsection{Effective Spin Model}\label{sec:spin_model}

At energies below the charge-transfer gap and the internal excitation energy of the localized ZR singlet, each doped plaquette can be represented by a local defect in the Cu-spin network. The central Cu moment is frozen due to the presence of the localized ZR singlet, and is then removed from the low-energy manifold. Nevertheless, the four surrounding Cu spins remain active and are coupled by the induced $J_2$ and $J_3$ exchanges. 

On a parent square lattice of $N=L^2$ Cu sites, we retain the spin-$1/2$ moments $\mathbf S_i$ on the intact Cu sites and represent each localized ZR singlet as a vcancy. The effective spin Hamiltonian for a given disorder configuration is
\begin{equation}
    H = J_1 \sum_{\langle ij \rangle_1}' \mathbf{S}_i \cdot \mathbf{S}_j + J_2 \sum_{\langle i j \rangle_2}^{\text{ZR}} \mathbf{S}_i \cdot \mathbf{S}_j
    + J_3 \sum_{ \langle i j \rangle_3 }^{\text{ZR}} \mathbf{S}_i \cdot \mathbf{S}_j.
    \label{eq:eff_H}
\end{equation}
Here $\langle ij\rangle_n$ denotes an $n$th-neighbor bond. The prime restricts the first sum to intact NN bonds, excluding the four $J_1$ bonds incident on every dopant center. The second sum with superscript ZR contains the four 2nd-NN pairs formed by Cu sites on adjacent sides of each ZR center, whereas the third sum contains the two 3rd-NN pairs located on opposite sides of each ZR center. These local $J_2$ and $J_3$ bond patterns are illustrated in the right inset of Fig.~\ref{fig:fig1}(a).

Since the $J_2$ and $J_3$ bonds connect sites on the same sublattice of the parent N\'eel state, each localized ZR center introduces a local frustrated unit into an otherwise bipartite Heisenberg antiferromagnet. In this sense, a hole-doped ZR center is represented not by a bare vacancy, but by a vacancy dressed with a specific pattern of frustrating exchange bonds.

For a hole concentration $\delta>0$, we sample $N_h=\delta N$ ZR centers from an uncorrelated spatial distribution and treat their positions as quenched disorder. We focus on the dilute regime, in which overlaps between the exchange patterns of nearby ZR centers are forbiden. The couplings $J_2$ and $J_3$ are taken to be uniform on all activated bonds, so that the disorder enters through the spatial arrangement of the defects rather than through random exchange amplitudes.

On the electron-doped side, the localized carrier removes the corresponding Cu moment and its incident $J_1$ bonds but does not activate the ZR-mediated $J_2$ and $J_3$ terms. The model then reduces to the conventional site-diluted AFM Heisenberg model~\cite{Keimer1992} illustrated in the left inset of Fig.~\ref{fig:fig1}(a). The distinction between these two defect structures, i.e., a bare vacancy for electron doping and a frustration-dressed vacancy for hole doping, defines the microscopic electron--hole asymmetry investigated below.

\section{Asymmetric Collapse of N\'eel Order}\label{sec:ground_state}

Having established the microscopic mechanism generating the $J_2$ and $J_3$ exchanges attributed to dopants, we now examine their macroscopic consequences on the ground-state magnetic phase diagram. To model the lightly doped regime, we consider a representative parameter set $J_2/J_1 = 0.5$ and $J_3/J_1 = 1.25$. These values are qualitatively consistent with the microscopic exchange hierarchy derived from the three-band Emery model in Sec.~\ref{sec:microscopic_model}, capturing the key feature that $J_3$ is the dominant frustrating component.
To accurately treat strong quantum fluctuations, spatial inhomogeneities, and magnetic frustration on an equal footing, we employ large-scale variational Monte Carlo (VMC) simulations on two-dimensional square lattices up to $L=20$, complemented by density matrix renormalization group (DMRG)~\cite{White1992,White1993} calculations on finite clusters. 

Within the VMC framework, physical spin operators are represented using the fermionic parton formulation, $\mathbf{S}_i = \frac{1}{2} f_{i\alpha}^\dagger \boldsymbol{\sigma}_{\alpha\beta} f_{i\beta}$, subject to the local single-occupancy constraint $\sum_\alpha f_{i\alpha}^\dagger f_{i\alpha} = 1$.  The trial wave function is then a Gutzwiller-projected state, $|\Psi\rangle = \mathcal{P}_G|\psi_0\rangle$. Here, the mean-field ansatz $|\psi_0\rangle$ is the ground state of a quadratic parton Hamiltonian, and the Gutzwiller projector $\mathcal{P}_G$ strictly enforces single occupancy, projecting $|\psi_0\rangle$ back into the physical spin Hilbert space. To capture local spatial inhomogeneities and frustration, the mean-field Hamiltonian incorporates site-dependent local Zeeman fields $\mathbf{h}_i (f_{i\uparrow}^\dagger f_{i\uparrow}-f_{i\downarrow}^\dagger f_{i\downarrow})/2$ as well as $d$-wave pairing terms; see the explict form of the mean-field Hamiltonian in Appendix~\ref{sec:vmc_appendix}. These variational parameters are optimized using the stochastic reconfiguration method~\cite{Sorella2001, Sorella2007}. Benchmarking against DMRG calculations confirms that this flexible ansatz faithfully captures the ground-state physics; see Appendix~\ref{sec:vmc_appendix} for details.

\subsection{Staggered Magnetization and Spin Structure Factor}\label{sec:Mz_Sq}
The long-range AFM order is quantified by the staggered magnetization $M_z$,
\begin{equation}
M_z = \frac{1}{N}\langle S^z(\mathbf{Q})\rangle, \quad \text{with}\quad S^a(\mathbf{q})=\sum_{i} e^{i \mathbf{q} \cdot \mathbf{r}_{i}} S^a_{i}, 
\label{eq:Mz}
\end{equation}
where $\mathbf{r}_{i}$ denotes the spatial coordinate of site $i$ and $\mathbf{Q}=(\pi, \pi)$ is the N\'eel ordering vector. To further map out magnetic correlations across the entire momentum space, we compute the static spin structure factor
\begin{equation}
\mathbb{S}(\mathbf{q})=\frac{1}{N^2}\sum_{a=x,y,z} \langle S^a(\mathbf{q})S^a(-\mathbf{q})\rangle.
\label{eq:Sq}
\end{equation}
For a state exhibiting long-range AFM order, the staggered magnetization $M_z$ is quantitatively consistent with $\sqrt{\mathbb{S}(\mathbf{Q})}$.

The ground-state staggered magnetization $M_z$ is computed as a function of doping concentration $\delta$ for both electron and hole doping. As shown in Fig.~\ref{fig:fig1}(a), it reveals a striking electron-hole asymmetric collapse of AFM N\'eel order. On the electron-doped side ($\delta < 0$), the magnetization $M_z$ decreases slowly. This behavior is consistent with the ``bare vacancy'' picture, where doped carriers effectively act as non-magnetic site dilutions. As noted in early neutron scattering studies on electron-doped cuprates~\cite{Keimer1992,Matsuda1992}, such simple site dilution can sustain long-range AFM order up to high doping levels, in sharp contrast to the hole-doped case~\cite{Luke1990,Keimer1992}.
This slow suppression of magnetization is rooted in the high classical percolation threshold of a two-dimensional square lattice ($p_c \approx 0.41$)~\cite{StaufferAharony1994,Sandvik2002}, which serves as an upper bound for the robustness of the AFM order against site dilution in the classical limit.
In stark contrast, hole doping ($\delta > 0$) causes a rapid and severe suppression of the AFM order, driving $M_z$ to vanish near a critical concentration $\delta_c \approx 0.08$. This expedited magnetic collapse is a direct macroscopic manifestation of the ``dressed'' ZR vacancies. Rather than passively diluting the lattice, each localized hole activates frustrating 2nd-NN $J_2$ and 3rd-NN $J_3$ superexchanges, which strongly disrupt the background N\'eel alignment far below the geometric percolation threshold. We find that the critical doping $\delta_c$ is only weakly dependent on the precise values of $J_2$ and $J_3$, as long as the hierarchy $J_3 \gtrsim J_1$ is maintained, indicating that the rapid suppression of AFM order on the hole-doped side is a generic feature.

The momentum-space signature of this frustrated magnetic destruction is directly captured by the static spin structure factor $\mathbb{S}(\mathbf{q})$ in Figs.~\ref{fig:fig1}(b--e). For the electron-doped state S1 ($\delta=-0.06$), the system still preserves a sharp Bragg peak at the ordering vector $\mathbf{Q}=(\pi, \pi)$; see Fig.~\ref{fig:fig1}(b). Under hole doping, even at a lower nominal concentration of $\delta=0.045$, the peak intensity in Fig.~\ref{fig:fig1}(c) is heavily suppressed and visibly broadened. As hole doping progresses to S3 ($\delta=0.075$), the commensurate peak near $\mathbf{Q}$ collapses, while spectral weight redistributes into emerging incommensurate features, as demonstrated in Fig.~\ref{fig:fig1}(d). Finally, at state S4 with $\delta=0.105$, the structure factor melts into a continuous ring-like diffuse pattern centered around $\mathbf{Q}$; see Fig.~\ref{fig:fig1}(e), indicating the complete destruction of long-range AFM order.

We emphasize that our numerical results are consistent with neutron-scattering experiments on both the electron- and hole-doped sides of the cuprate phase diagram.
On the electron-doped side, the magnetic response remains predominantly commensurate and centered at the N\'eel wave vector $\mathbf{Q}=(\pi,\pi)$ across a broad range of doping and energy~\cite{Yamada2003,Fujita2008,FUJITA2003130,Wilson2006,Fujita2011}, consistent with the sharp peak retained in Fig.~\ref{fig:fig1}(b). By contrast, lightly hole-doped cuprates develop pronounced incommensurate magnetic correlations in the insulating spin-glass regime~\cite{Matsuda2000PRB,Fujita2002}. This is in qualitative agreement with the emergence of incommensurate spectral weight away from $\mathbf{Q}$ in Figs.~\ref{fig:fig1}(d,e).

\subsection{Emergent Domain-Wall Structure}\label{sec:domain_wall}

\begin{figure}[t]
  \centering
  \includegraphics[width=0.48\textwidth]{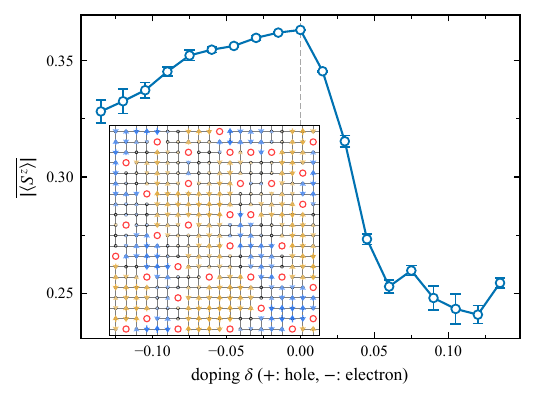}
\caption{Average magnitude of the local spin moments, $\longbar{|\langle S^z\rangle|}$, as a function of signed doping concentration $\delta$ (+: hole, $-$: electron) on an $L=20$ lattice obtained from VMC. Inset: Real-space distribution of the local staggered magnetization $m^z_i$ obtained by VMC. Blue and yellow arrows denote local magnetic moments in opposite AFM domains. The red open circles represent the doped holes (vacancies). Here we use $L=20$, $J_2$=0.5, $J_3$=1.25, and $\delta=0.09$.}\label{fig:domain}
\end{figure}

\begin{figure*}[t]
  \centering
  \includegraphics[width=0.92\textwidth]{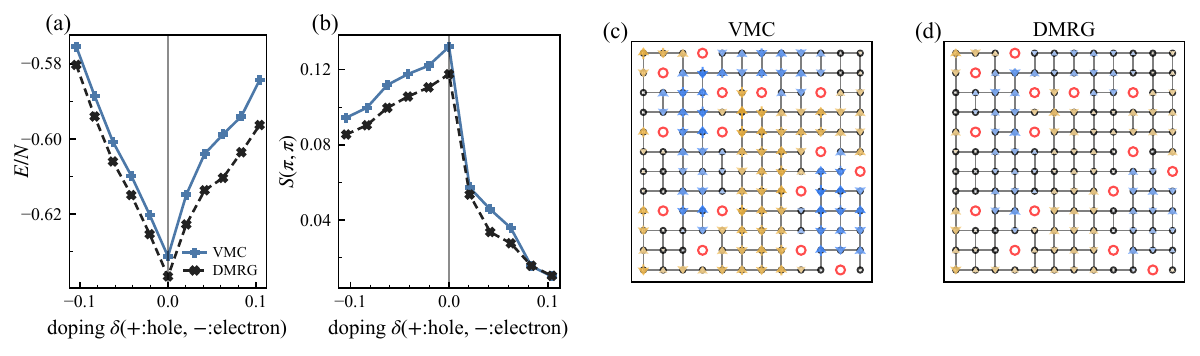}
\caption{Benchmark between VMC and DMRG results on the $12\times12$ lattice with open boundary conditions. 
(a) Energy per active site, $E/N$, as a function of signed doping $\delta$, where hole doping is plotted on the positive axis and electron doping on the negative axis. Here $N=12\times12-N_h$ excludes the defect sites. $J_1=1.0$, $J_2=0.5$, $J_3=1.25$. 
(b) Spin structure factor $\mathbb{S}(\pi,\pi)$ as a function of signed doping.
(c,d) Domain-wall structure in real-space $\langle S^z_i\rangle$ distributions for hole doping with $N_h=15$, obtained from VMC and DMRG, respectively. Blue and yellow arrows denote local magnetic moments in opposite AFM domains. Red circles mark the defect sites. The error bars of VMC calculations are comparable or smaller than the marker size.  DMRG calculations utilize a bond dimension up to $D=10000$ under $\mathrm{U}(1)$ spin symmetry, with truncation errors below $1 \times 10^{-5}$. }~\label{fig:benchmark}
\end{figure*}

The frustration introduced by the dopant-mediated $J_2$ and $J_3$ exchanges raises an intriguing question regarding whether the ground state could realize a quantum spin liquid. To systematically address this, we examine several candidate variational mean-field ansatzes within the VMC framework, including coplanar magnetic states (where local moments $\mathbf{m}_i$ lie in a plane) and noncoplanar spin chirality states. Through rigorous energy optimization, we find that the collinear magnetic ansatz consistently achieves the lowest variational energy across the lightly doped regime.

To elucidate the physical nature of the ground state, we evaluate the spatial average of the local spin moment magnitude, 
\begin{equation}
\longbar{|\langle S^z\rangle|} = \frac{1}{N}\sum_i |\langle S^z_i \rangle|,
\end{equation}
as plotted in Fig.~\ref{fig:domain}. While the global N\'eel order parameter $M_z$ collapses abruptly on the hole-doped side ($\delta > 0.08$), the local spin moment magnitude remains strikingly robust ($\longbar{|\langle S^z\rangle|} \sim 0.25$--$0.3$). This substantial local moment explicitly disfavors a quantum spin liquid ground state with quenched local spins; rather, it demonstrates that the suppression of macroscopic AFM order is driven by spatial phase disordering.

The microscopic origin of this phase disordering is straightforwardly revealed by the real-space distribution of $\langle S^z_i\rangle$, as illustrated in the inset of Fig.~\ref{fig:domain} for a representative hole-doped configuration with $\delta = 0.09$. The spatial landscape fragments into well-defined domains with opposite AFM order parameters ($\pi$ phase-shifted), visually resembling a black-and-white Go board. The exact geometric configuration and boundary contours of these domains are strictly dictated by the spatial distribution of the localized ZR centers. Within the bulk of each domain, local spins remain strongly aligned. However, at the domain-wall boundaries separating adjacent domains, the local variational parameters $m^z_i$ drop significantly. The system thus retains a short-range magnetic solid, where the emergence of domain walls destroys global spatial phase coherence.

This domain-wall structure provides a direct physical mechanism for the momentum-space patterns observed in $\mathbb{S}(\mathbf{q})$; see Figs.~\ref{fig:fig1}(c--e). Within the doping regime where domain-wall structure emerges, each domain wall forces a $\pi$ phase flip in the local AFM order, partitioning the lattice into regions with a characteristic scale $D$. This spatial phase modulation introduces an additional wavevector, which shifts the original Bragg peak at $\mathbf{Q}=(\pi, \pi)$ to incommensurate wavevectors $\mathbf{Q} \pm \boldsymbol{\delta q}$ with a displacement magnitude $\delta q \sim \pi/D$. Furthermore, upon ensemble averaging over random dopant configurations, the irregular domain orientations and size fluctuations broaden and smooth these incommensurate features into the continuous ring-like diffuse pattern centered at $(\pi, \pi)$, as shown in Fig.~\ref{fig:fig1}(e).

\subsection{Quantitative Benchmarks via DMRG}\label{sec:benchmark}
To further establish the physical picture of the asymmetric phase diagram using an unbiased numerical method, we perform a quantitative comparison between DMRG and VMC calculations on a $12\times 12$ square lattice with open boundary conditions (OBC). The DMRG results are converged using a bond dimension up to $D=10000$, ensuring a truncation error below $1 \times 10^{-5}$. Because the exact $\mathrm{SU}(2)$ spin rotational symmetry of Eq.~\eqref{eq:eff_H} precludes spontaneous symmetry breaking on finite clusters~\cite{Stoudenmire2012}, a local pinning field $H_{\text{pinning}} = \sum_{i\in\text{corners}}(-1)^{x_i+y_i}B S^z_i$ (with $B=0.3$) is applied at the lattice corners in both methods. This explicitly breaks the $\mathrm{SU}(2)$ symmetry and stabilizes the AFM order for a direct comparison of local observables.

We find that VMC and DMRG calculations exhibit notable agreement on the doping-dependent evolution of the AFM order,  as illustrated in Fig.~\ref{fig:fig1}(a). While the absolute magnitudes of $M_z$ show minor discrepancies due to finite-size and boundary effects, the VMC preciously results reproduce the doping-dependent trend obtained by DMRG. This quantitative agreement firmly establishes the reliability of our large-scale ($L=20$) VMC calculations for mapping out the asymmetric phase diagram.

The detailed benchmark results, summarized in Fig.~\ref{fig:benchmark}, demonstrate multi-level quantitative consistency between DMRG and VMC across global energetics, macroscopic order parameters, and microscopic real-space spin textures. 
As shown in Fig.~\ref{fig:benchmark}(a), the per-active-site $E/N$ energy obtained from VMC matches the DMRG data across both electron- and hole-doped regimes, confirming that the variational ansatz captures the ground-state energetics with high precision. Moreover, the spin structure factor $\mathbb{S}(\mathbf{Q})$ in Fig.~\ref{fig:benchmark}(b) calculated by both methods independently confirms the magnetic asymmetry, capturing the slow decay under electron doping and the rapid collapse under hole doping without discrepancy.

Crucially, this quantitative agreement extends down to the microscopic real-space level. Figures~\ref{fig:benchmark}(c) and \ref{fig:benchmark}(d) display the local spin moment distributions $\langle S^z_i\rangle$ for an identical random defect configuration with $N_h=15$. VMC and DMRG yield virtually identical real-space spin textures, matching local moment magnitudes site by site across the entire lattice and reproducing the exact same domain-wall contours. This microscopic agreement confirms that the fragmented domain-wall structure is an intrinsic physical property of the model rather than a method-dependent artifact.
This comprehensive multi-level consistency between unbiased DMRG and large-scale VMC demonstrates that the dopant-driven magnetic asymmetry and frustration-induced N\'eel collapse are intrinsic physical features of the model, fully independent of the numerical method employed.


\section{Finite-Temperature Magnetism of the 3D Layered Model}\label{sec:finite_T}

\begin{figure}[t]
  \centering
  \includegraphics[width=0.48\textwidth]{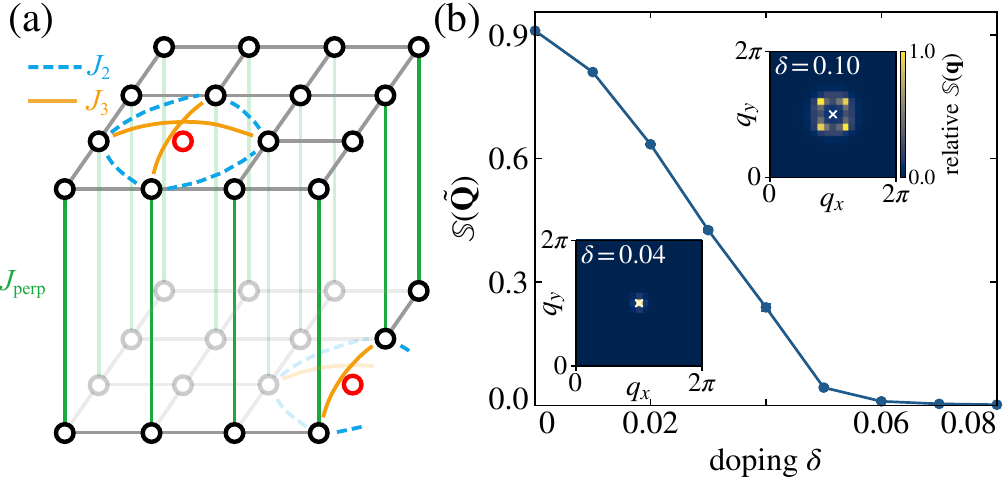}
\caption{
Schematic of the 3D layered spin model and magnetic response under hole doping. (a) Schematic representation of the 3D layered lattice with weak interlayer exchange $J_\perp$. Each 2D layer contains randomly distributed localized Zhang--Rice singlets (open red circles) that activate frustrating intralayer 2nd-NN $J_2$ (blue dashed lines) and 3rd-NN $J_3$ (orange arrows) exchanges between intact Cu spins. 
(b) 3D static spin structure factor $\mathbb{S}(\tilde{\mathbf{Q}})$ at the 3D N\'eel ordering vector $\tilde{\mathbf{Q}}=(\pi, \pi, \pi)$ as a function of hole doping concentration $\delta$.
Inset: The relative intensity of $\mathbb{S}(\mathbf{q})$ in the $q_z=\pi$ plane with $\delta = 0.04$ (left) and $\delta=1.0$ (right). The exchange parameters are set to $J_1=1.0$, $J_2=0.5$, $J_3=1.25$, and $J_\perp=0.1$. The temperature is fixed at $T=0.1$ for all calculations.
}
\label{fig:3D_model}
\end{figure}

To bridge our ground-state analysis with the experimental phase diagram~\cite{Armitage2010,Keimer1992, Kastner1998}, a central task is to evaluate the doping dependence of the N\'eel transition temperature $T_N$ and explore potential non-N\'eel states at finite temperatures. 
In a purely two-dimensional system with continuous $\mathrm{O}(3)$ spin symmetry, the Mermin-Wagner theorem strictly precludes spontaneous symmetry breaking at any non-zero temperature ($T > 0$).
In real bulk crystals, however, a finite $T_N$ is stabilized by the weak interlayer exchange coupling $J_\perp$ that connects adjacent CuO$_2$ planes.

To capture this 3D physics, we extend our 2D effective spin Hamiltonian in Eq.~\eqref{eq:eff_H} into a 3D layered geometry, as schematically depicted in Fig.~\ref{fig:3D_model}(a). Treating the spin-$1/2$ operators as classical $\mathrm{O}(3)$ unit vectors, the 3D Hamiltonian defined on an $L \times L \times L$ cubic lattice is given by
\begin{equation}
    H_{\text{3D}} = \sum_{m=1}^{L} H_{\text{2D}}^{(m)} + J_\perp \sum_{\langle m,n\rangle} \sum_{i}' \mathbf{S}_{i,m} \cdot \mathbf{S}_{i,n},
    \label{eq:H_3D}
\end{equation}
where $\mathbf{S}_{i,m}$ denotes the classical spin vector at site $i$ in layer $m$, and $H_{\text{2D}}^{(m)}$ represents the classical counterpart of Eq.~\eqref{eq:eff_H} for the $m$-th plane. The second term introduces a weak interlayer exchange $J_\perp = 0.1 J_1$ between vertically adjacent intact Cu spins. Note that the primed sum $\sum_i'$ excludes vertical bonds incident on ZR-center vacancies; see Fig.~\ref{fig:3D_model}(a). To model quenched disorder in bulk single crystals, each layer maintains the same nominal doping concentration $\delta$, while the spatial locations of localized ZR centers are generated independently across different layers.


As a direct starting point, we examine the static spin structure factor $\mathbb{S}(\mathbf{q})$ of the 3D model at a low temperature ($T=0.1 J_1$). Because the quenched disorder and emergent frustration give rise to a complex energy landscape, the simulations are performed using classical Monte Carlo enhanced by the parallel tempering algorithm~\cite{Hukushima1996,Earl2005} to efficiently overcome slow thermal relaxation. As displayed in Fig.~\ref{fig:3D_model}(b), the 3D Bragg peak intensity $\mathbb{S}(\tilde{\mathbf{Q}})$ at the N\'eel vector $\tilde{\mathbf{Q}}=(\pi,\pi,\pi)$ collapses abruptly upon hole doping. Simultaneously, the in-plane spectral weight melts into a diffuse ring-like pattern centered at $(\pi,\pi)$ [inset of Fig.~\ref{fig:3D_model}(b)]. Remarkably, this finite-temperature spectral feature directly mirrors the 2D quantum ground-state structure factor obtained via VMC in Fig.~\ref{fig:fig1}(e), confirming that the phase-disordered magnetic state persists robustly into the 3D bulk stack. This rapid breakdown of 3D coherent magnetic scattering motivates a systematic evaluation of the finite-temperature phase boundaries.

\begin{figure}[t]
  \centering
  \includegraphics[width=0.48\textwidth]{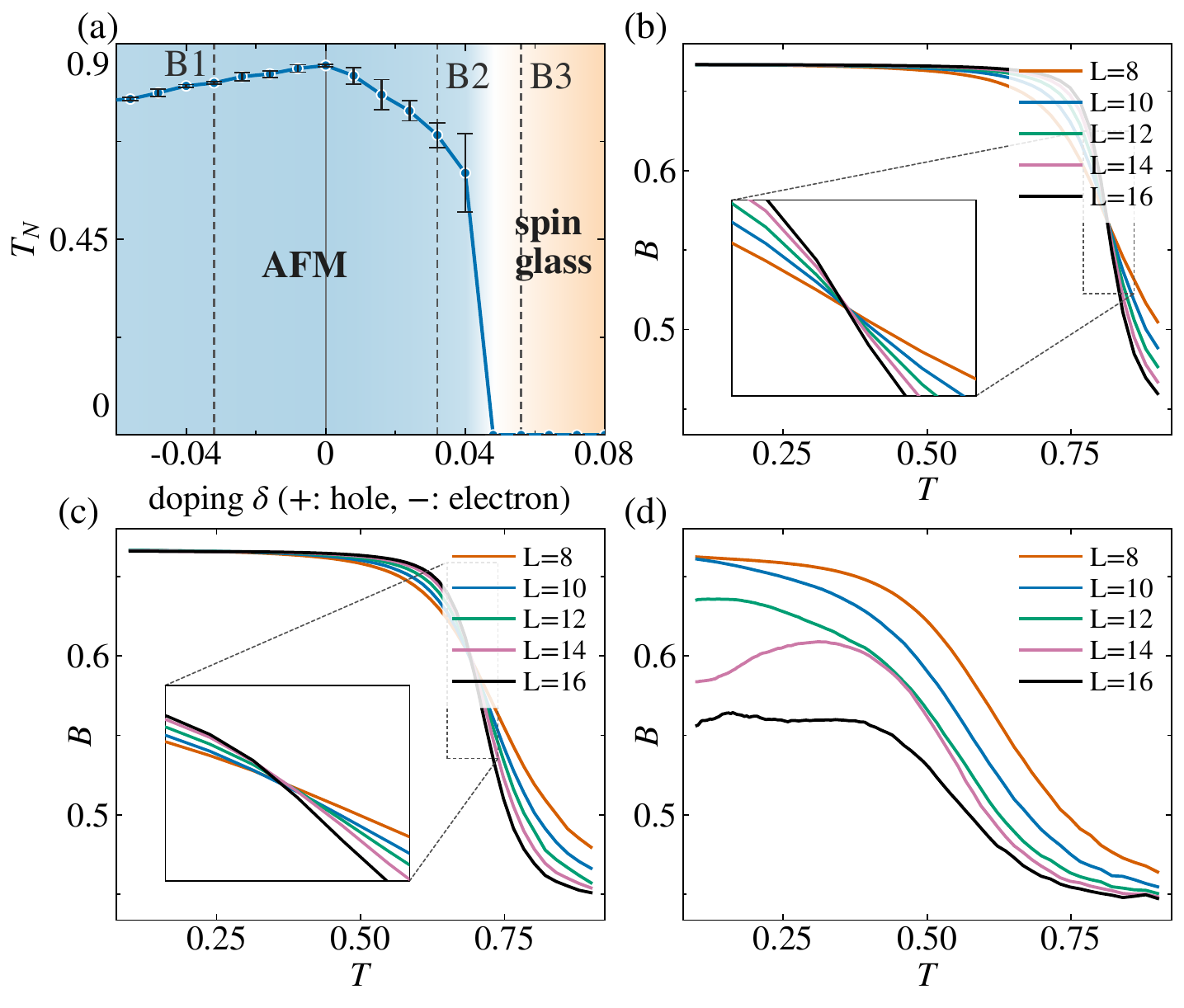}
\caption{Finite-temperature phase diagram obtained from classical Monte Carlo simulations of Eq.~(\ref{eq:H_3D}). (a) Extracted N\'eel temperature $T_N$ as a function of doping. 
(b)-(d) Temperature dependence of the Binder ratio $B(T)$ for system sizes ranging from $L=8$ to $16$ at three representative doping levels, indicated by the vertical dashed lines B1, B2, and B3 in (a). The insets provide magnified views of the curves to clearly resolve the presence [(b), (c)] or absence [(d)] of a crossing point. The simulations are performed on a 3D $L \times L \times L$ lattice with classical O(3) spins using $J_1=1$, $J_2=0.5$, $J_3=1.25$, and $J_\perp=0.1$.}~\label{fig:fig2}
\end{figure}

\subsection{Doping Dependence of the N\'eel Temperature $T_N$}\label{sec:TN}

To determine the critical phase boundary $T_N(\delta)$, we implement the 3D staggered AFM magnetization defined as
\begin{equation}
M_z = \frac{1}{N_{\text{tot}}} \sum_{i,m} e^{i \tilde{\mathbf{Q}} \cdot \mathbf{r}_{i,m}} S^z_{i,m},
\label{eq:Mz_3D}
\end{equation}
where $\mathbf{r}_{i,m}$ denotes the 3D spatial coordinate of site $i$ in layer $m$, and $N_{\text{tot}}$ is the total number of active Cu sites. 
However, the staggered magnetization does not show a sharp onset in finite systems. Therefore, the phase boundary cannot be reliably determined from $M_z$ alone. To accurately extract the N\'eel temperature $T_N$, we instead compute the Binder ratio~\cite{Binder1981},
\begin{equation}
B(T) = \langle M_z^4 \rangle_T / \langle M_z^2 \rangle^2_T,
\end{equation}
where $\langle \dots \rangle_T$ represents the average over both the thermal ensemble and the random vacancy configurations. Since the dimensionless quantity $B(T)$ is scale-invariant at a continuous phase transition~\cite{Binder1981}, its value becomes independent of the system size $L$ at the critical point, e.g., $T_N$. Therefore, $T_N$ can be precisely determined by the crossing point of the $B(T)$ curves for different $L$, as demonstrated in Figs.~\ref{fig:fig2}(b) and \ref{fig:fig2}(c).

The resulting finite-temperature phase diagram is presented in Fig.~\ref{fig:fig2}(a). The asymmetric evolution of the classical $T_N$ curve is highly consistent with our 2D quantum ground-state calculations. For a given doping, $T_N$ is extracted from the crossing point of the Binder ratio curves $B(T)$, as shown for representative electron ($\delta = -0.032$) and hole ($\delta = 0.032$) dopings in Figs.~\ref{fig:fig2}(b) and \ref{fig:fig2}(c), respectively. On the electron-doped side, $T_N$ is only gradually suppressed, reflecting a classical site-dilution percolation process. Conversely, on the hole-doped side, the frustrating intralayer $J_2$ and $J_3$ couplings efficiently disrupt long-range correlations, leading to a rapid suppression of $T_N$. At higher hole dopings, such as $\delta = 0.048$, the curves of $B(T)$ no longer cross, indicating the vanishing of $T_N$ and the absence of long-range AFM order; see Fig.~\ref{fig:fig2}(d). The AFM phase thus terminates at a critical hole doping of $\delta_c \approx 0.05$, which closely matches experimental values for cuprates~\cite{Armitage2010,Keimer1992}.

We note that the critical hole doping $\delta_c \approx 0.05$ is insensitive to the precise value of $J_\perp$, which primarily determines the magnitude of $T_N$ for a given doping. This distinction can be naturally understood within the random phase approximation (RPA) framework for quasi-2D antiferromagnets~\cite{Scalapino1975,Irkhin1999,Yasuda2005}. Within RPA, the 3D transition temperature is governed by $$T_N \sim  \frac{\rho_s}{\ln(A J_1 / J_\perp)},$$ where $\rho_s$ denotes the 2D in-plane spin stiffness and $A$ is a constant. While $J_\perp$ sets the thermal scale $T_N$ by facilitating interlayer phase locking, the ultimate collapse at $\delta_c$ is attributed mainly to the internal destruction of the in-plane magnetic stiffness ($\rho_s \to 0$) driven by the dopant-induced $J_2$--$J_3$ frustration.

\subsection{Emergence of the Spin-Glass Phase}\label{sec:SG}

Given the rapid suppression of the long-range AFM order upon hole doping, a natural question arises: what magnetic state takes over at finite temperatures once the N\'eel order is destroyed? To address this, we investigate the potential spin-glass (SG) phase using parallel tempering Monte Carlo simulations~\cite{Hukushima1996,Earl2005}.
Following standard practice for classical spin glasses~\cite{Lee2003}, we introduce two independent replicas, $\{\mathbf{S}_{i,m}^{(1)}\}$ and $\{\mathbf{S}_{i,m}^{(2)}\}$, sharing identical vacancy configurations but evolving independently with two identical thermal baths.

\begin{figure}[t]
  \centering
  \includegraphics[width=0.38\textwidth]{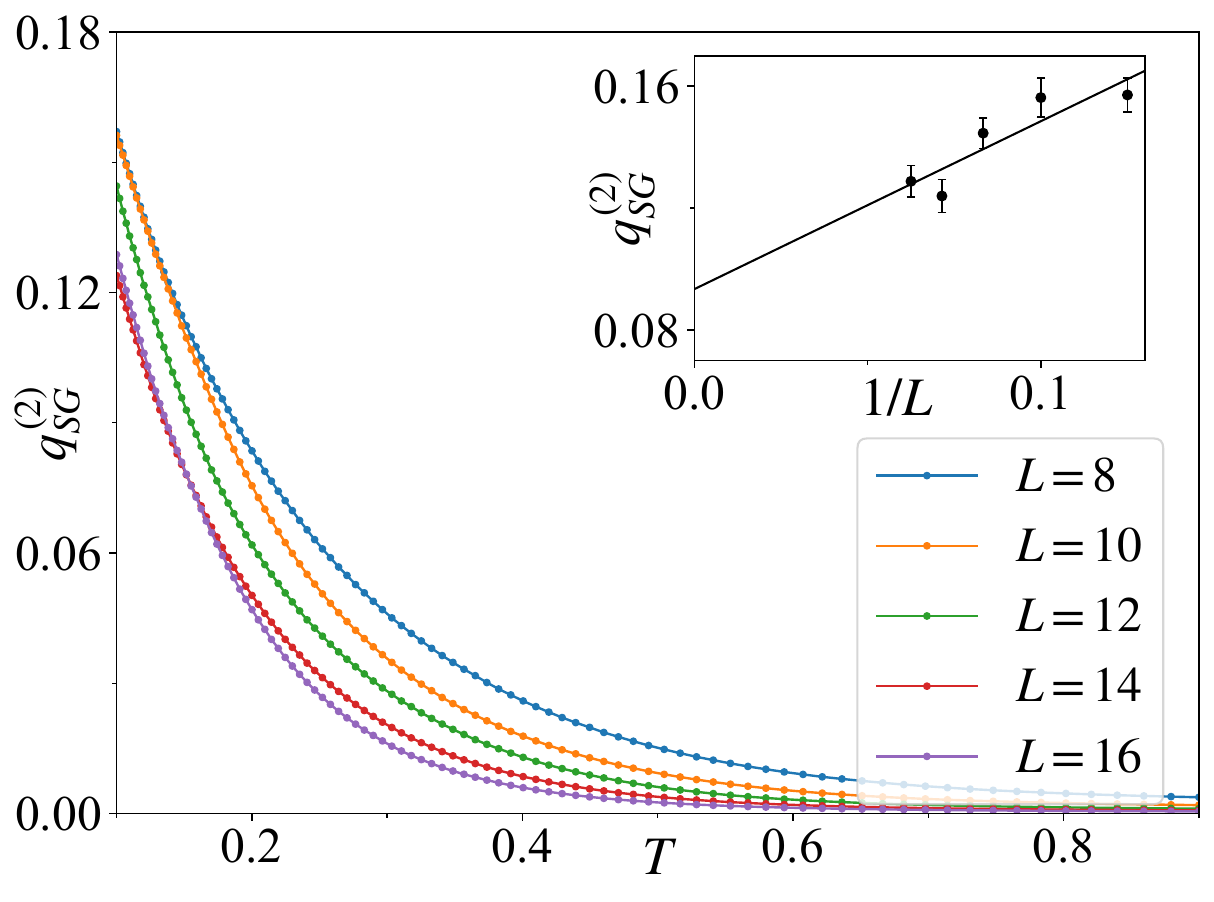}
\caption{Emergence of the spin-glass phase. 
Temperature dependence of the Edwards-Anderson spin-glass order parameter $q^{(2)}_{SG}$ for various system sizes $L$. Inset: The finite-size scaling of $q^{(2)}_{SG}$ versus $1/L$ at low temperature $T=0.1$. The non-zero intercept in the thermodynamic limit ($L \to \infty$) provides direct numerical evidence for a robust spin-glass phase. 
}~\label{fig:fig3}
\end{figure}

The macroscopic 3D spin-glass order is quantified by the squared Edwards-Anderson order parameter~\cite{Binder1986},
\begin{equation}
q_{\text{SG}}^{(2)} = \left\langle \left( \frac{1}{N} \sum_{i,m} \mathbf{S}_{i,m}^{(1)} \cdot \mathbf{S}_{i,m}^{(2)} \right)^2 \right\rangle_{\!T}.
\label{eq:q_SG}
\end{equation}
We consider a representative hole doping $\delta = 0.09$, where the AFM order paramter $\mathbb{S}(\tilde{\mathbf{Q}})$ is completely destroyed.As shown in Fig.~\ref{fig:fig3}, $q^{(2)}_{SG}$ exhibits a sharp onset at low temperatures. Furthermore, finite-size scaling of $q^{(2)}_{SG}$ versus $1/L$ at $T=0.1$ yields a non-zero intercept as $L \to \infty$.
The SG phase is also supported by the momentum-space signatures. For a single realization of quenched dopants, the low-temperature spin structure factor $\mathbb{S}(\mathbf{q})$ does not feature a single N\'eel peak. Instead, it displays several incommensurate spots, whose positions are dictated by the particular spatial configuration of localized ZR centers.
Upon ensemble-average, these sample-specific peaks merge into a $C_4$-symmetric ring centered at $(\pi, \pi)$, with intensity maxima at four incommensurate wavevectors; see inset of Fig.~\ref{fig:3D_model}(b). This strong sample-to-sample fluctuation in momentum space, combined with the non-zero $q_{\text{SG}}^{(2)}$, confirms that the system freezes into a glass-like state.

The microscopic origin of this 3D spin-glass state is a direct consequence of the 2D domain-wall structures identified in Sec.~\ref{sec:microscopic_model}. In an unfrustrated quasi-2D antiferromagnet, a weak interlayer coupling $J_\perp$ easily overcomes thermal fluctuations to lock the global in-plane N\'eel order across layers, breaking continuous SU(2) symmetry, and establishing a finite-temperature AFM order~\cite{Scalapino1975,Yasuda2005}. On the hole-doped side, however, the intralayer $J_2$--$J_3$ frustration fragments each 2D plane into local rigid domains separated by static domain walls with $\pi$ phase shifts. Because the spatial locations of localized ZR centers vary randomly from layer to layer, the domain-wall patterns in adjacent planes are spatially uncorrelated.

This interlayer spatial mismatch prevents $J_\perp$ from establishing coherent global phase locking across the 3D stack. Instead, $J_\perp$ generates competing local alignments between random in-plane domains. 
This inevitable spatial mismatch of domain walls completely prevents any global alignment, trapping the system in a highly degenerate energy landscape. Consequently, the system freezes into a 3D SG at finite temperatures~\cite{Saunders2007,Andreanov2010}.
Consequently, as thermal energy is lowered, the system freezes into a robust 3D spin glass, providing a unified microscopic origin for the spin-glass phase universally observed in lightly hole-doped cuprates~\cite{Chou1993,Niedermayer1998,Kastner1998,Julien_2003,Matsuda2000PRB}. We emphasize that because electron dopants act as simple non-magnetic vacancies without generating intralayer frustration or domain walls, no spin-glass phase emerges on the electron-doped side, consistent with both theoretical models and experimental phase diagrams~\cite{Armitage2010,Keimer1992}.

\section{Conclusion}\label{sec:discussion}

In summary, we have proposed a unified microscopic framework that resolves the doping-driven magnetic transitions and electron-hole asymmetry in cuprates. By downfolding the three-band Emery model in the presence of realistic local dopant potential $E_{loc}$, we demonstrate that spatially localized ZR singlets do not act merely as inert non-magnetic vacancies. Instead, they serve as active intermediate quantum states for virtual charge fluctuations, mediating emergent the 2nd NN $J_2$ and 3rd NN $J_3$ exchanges across the dopant plaquette. This active exchange pathway fundamentally revises the conventional spin-dilution paradigm, establishing that local dopant potentials generate substantial intralayer magnetic frustration.

Through a combination of large-scale 2D quantum simulations (VMC and DMRG) and 3D classical Monte Carlo calculations, we systematically map out the macroscopic consequences of this dopant-induced frustration. In 2D, the emergent $J_2$--$J_3$ couplings fragment the Néel background into static domain-wall structures, driving a rapid collapse of long-range AFM order in the hole-doping side. In 3D layered stacks, the spatial mismatch of these in-plane domain walls prevents global phase locking, causing $T_N$ to drop sharply and forcing the system to freeze into a robust 3D spin glass phase upon hole doping. In sharp contrast, electron doping lacks these frustrating channels, reducing to simple classical site dilution where AFM order persists up to high doping levels. Our results thus identify local dopant potentials as the microscopic origin of the long-standing magnetic electron–hole asymmetry, establishing localized ZR physics as an indispensable missing ingredient in understanding the lightly doped regime of cuprates.

For future study, we propose that the emergent frustrating spin textures and domain-wall boundaries surrounding individual localized dopants can be directly mapped in real space using spin-polarized scanning tunneling microscopy (SP-STM)~\cite{Wiesendanger2009}.
In momentum space, the predicted melting of the $(\pi,\pi)$ Bragg peak into a $C_4$-symmetric diffuse pattern qualitatively aligns with neutron scattering observations in lightly hole-doped cuprates~\cite{Fujita2002,Matsuda2000PRB}, where incommensurate features are preferentially oriented along the $45^\circ$ diagonal direction~\cite{Fujita2002}. This directional selection suggests that inter-dopant interactions may play a vital role in orienting the domain walls, offering a promising avenue for future studies.
Beyond cuprates, the concept of dopant-mediated emergent exchange provides a general mechanism for understanding carrier localization and frustrating interactions in a wide spectrum of lightly doped strongly correlated materials.

\begin{acknowledgments}
We acknowledge the support by the National Natural Science Foundation of China (Grant NSFC-12494594, NSFC-12504180, NSFC-12274290, NSFC-12347107, NSFC-12474146, and NSFC-12522406), the Ministry of Science and Technology  (Grant No. 2022YFA1405400), the Chinese Academy of Sciences Project for Young Scientists in Basic Research (2022YSBR-048), the Innovation Program for Quantum Science and Technology (2021ZD0301902), the New Cornerstone Investigator Program, and the Beijing Natural Science Foundation (No. JR25007). 
HKJ acknowledges the support from the start-up funding from ShanghaiTech University.
\end{acknowledgments}


\appendix

\section{Derivation of the Effective Spin Hamiltonian}~\label{app:downfolding}

This section provides the technical details of the perturbative downfolding summarized in Sec.~\ref{sec:microscopic_model}. We use the three-band Emery Hamiltonian in Eq.~\eqref{eq:three_band_main}, together with the hole-representation and hopping-sign conventions defined there.

\subsection{Undoped Superexchange $J_1$}

\begin{figure}[!h]
  \centering
  \includegraphics[width=\columnwidth]{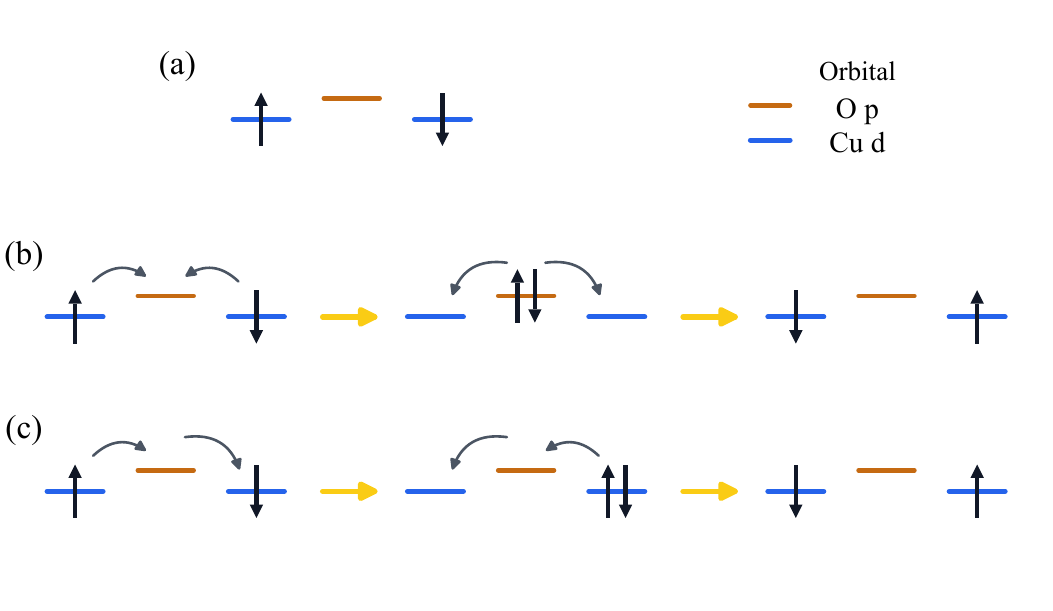}
\caption{Schematic illustration of the $J_1$ superexchange process.
(a) Three-site Cu--O--Cu geometry used for the $J_1$ path, where blue and orange lines denote Cu $d$ and O $p$ orbitals, respectively.
In the hole representation, the O $p$ level has chemical potential $\xi_p>0$ relative to the Cu $d$ level, which is taken as the zero reference.
(b,c) Two representative virtual hopping channels contributing to $J_1$.
Gray curved arrows indicate hole hopping within each configuration, yellow arrows denote the evolution of virtual states, and vertical arrows represent hole spins.}~\label{fig:Appexdix1}
\end{figure}

To derive the first-nearest-neighbor superexchange $J_1$, we consider the minimal three-orbital Cu--O--Cu cluster shown in Fig.~\ref{fig:Appexdix1}(a), consisting of two adjacent Cu sites (labeled 1 and 2) bridged by a single O site. In the undoped limit, this cluster contains exactly two holes. Throughout this derivation, we work in the hole representation and take the Cu $d$ level as the zero of energy, while the O $p$ level lies at energy $\xi_p>0$, as illustrated in Fig.~\ref{fig:Appexdix1}(a). We emphasize that this $J_1$ process is entirely local and only involves the linear Cu--O--Cu bond.

We employ the block-matrix downfolding technique to obtain the effective low-energy Hamiltonian $H_{\text{eff}}$ within the subspace where each Cu site is occupied by exactly one hole. Using the resolvent expansion, the effective Hamiltonian is written as
\begin{equation}
    H_{\text{eff}}(\omega) = H_{00} + T_{01}(\omega - H_{11})^{-1}T_{10} + \dots
\end{equation}
and for the low-energy projection we set $\omega=0$.

For the $S^z=0$ sector, the Hilbert space is 9-dimensional. The basis vectors are explicitly chosen as
$\Psi_{\text{basis}} = ( d_{2\downarrow}^\dagger d_{1\uparrow}^\dagger, \ 
        d_{2\uparrow}^\dagger d_{1\downarrow}^\dagger, \ 
        d_{2\downarrow}^\dagger p_{\uparrow}^\dagger, \ 
        d_{2\uparrow}^\dagger p_{\downarrow}^\dagger, \ 
        p_{\downarrow}^\dagger d_{1\uparrow}^\dagger, \ 
        p_{\uparrow}^\dagger d_{1\downarrow}^\dagger, \ 
        d_{2\downarrow}^\dagger d_{2\uparrow}^\dagger, \  
        d_{1\downarrow}^\dagger d_{1\uparrow}^\dagger, \ \\ 
        p_{\downarrow}^\dagger p_{\uparrow}^\dagger ) |0 \rangle.$
In this basis, the Hamiltonian takes a block form 
\begin{equation*}
H_{\perp} = \begin{pmatrix} H_{00} & T_{01} & 0 \\ T_{10} & H_{11} & T_{12} \\ 0 & T_{21} & H_{22} \end{pmatrix}.
\end{equation*} 
We rewrite $H_{\perp}$ explicitly in terms of the 9-dimensional basis $\Psi_{\text{basis}}$ as
\begingroup
\setlength{\arraycolsep}{2pt}
\begin{equation*}
    H_{\perp} =
    \begin{pmatrix}
        0 & 0 & -t_{pd} & 0 & t_{pd} & 0 & 0 & 0 & 0  \\
        0 & 0 & 0 & -t_{pd} & 0 & t_{pd} & 0 & 0 & 0  \\
        -t_{pd} & 0 & \xi_p & 0 & 0 & 0 & t_{pd} & 0 & t_{pd} \\
        0 & -t_{pd} & 0 & \xi_p & 0 & 0 & -t_{pd} & 0 & -t_{pd} \\
        t_{pd} & 0 & 0 & 0 & \xi_p & 0 & 0 & -t_{pd} & -t_{pd} \\
        0 & t_{pd} & 0 & 0 & 0 & \xi_p & 0 & t_{pd} & t_{pd} \\
        0 & 0 & t_{pd} & -t_{pd} & 0 & 0 & U_d & 0 & 0 \\
        0 & 0 & 0 & 0 & -t_{pd} & t_{pd} & 0 & U_d & 0 \\
        0 & 0 & t_{pd} & -t_{pd} & -t_{pd} & t_{pd} & 0 & 0 & 2\xi_p+U_p \\
    \end{pmatrix}.
\end{equation*}
\endgroup
The low-energy subspace $H_{00}$ is spanned by the first two states, corresponding to one hole on each Cu site. Up to fourth order in $t_{pd}$, and assuming $U_d,\, 2\xi_p+U_p \gg t_{pd}$, the downfolded Hamiltonian becomes
\begin{align}
    H_{\text{eff}} &\approx -T_{01} H_{11}^{-1} \left( \mathbf{I} + T_{12} H_{22}^{-1} T_{21}H_{11}^{-1} \right) T_{10} \nonumber \\
    &= -\frac{2 t_{pd}^2}{\xi_p}\mathbf{I} -\frac{2t_{pd}^4}{\xi_p^2}\left(\frac{1}{U_d}+\frac{2}{2\xi_p +U_p}\right)
    \begin{pmatrix}
        1 & -1 \\
        -1 & 1
    \end{pmatrix}.
\end{align}
The first term is a spin-independent uniform second-order energy shift and therefore does not contribute to the singlet--triplet splitting. The second term arises from fourth-order virtual charge-transfer processes, illustrated schematically in Fig.~\ref{fig:Appexdix1}(b) and (c), and directly determines the splitting.
Mapping this result to the Heisenberg form
\begin{equation}
    H_{\text{spin}} = J_1\left(\mathbf{S}_1\cdot\mathbf{S}_2-\frac{1}{4}\right),
\end{equation}
we obtain the standard nearest-neighbor superexchange
\begin{equation}
    J_1 = \frac{4t_{pd}^4}{\xi_p^2}
    \left(\frac{1}{U_d}+\frac{2}{2\xi_p +U_p}\right)
    \approx
    \frac{4t_{pd}^4}{\xi_p^2}
    \left(\frac{1}{U_d}+\frac{1}{\xi_p}\right),
\end{equation}
where in the final step we set $U_p=0$, consistent with the numerical parameters used in the main text.

Importantly, this result is completely independent of the intrinsic alternating signs discussed previously. The virtual charge-transfer processes contributing to $J_1$ occur strictly along the linear Cu--O--Cu bond and do not form any closed loops involving O--O hopping $t_{pp}$. Consequently, any alternating signs or phase factors associated with the $t_{pd}$ matrix elements cancel exactly, since the perturbative contributions depend only on $|t_{pd}|^2$ and $|t_{pd}|^4$.

\subsection{Zhang-Rice Singlet Formulation}

The local CuO$_4$ geometry is shown in Fig.~\ref{fig:ZR}(a). The symmetry-adapted ligand orbitals, their energies, and the resulting low-energy ZR Hamiltonian are given in Eqs.~\eqref{eq:ligand_orbitals}--\eqref{eq:zr_local_hamiltonian}. Here we retain the full two-hole cluster Hamiltonian to show how the low-energy singlet sector used in the main text is obtained.

The ground state of the local cluster then contains two holes: one intrinsic hole on the Cu site and one doped hole occupying the surrounding oxygen orbitals. Restricting to the subspace spanned by the Cu $d$ orbital and the symmetry-matched ligand $b$ orbital, the two-hole Hamiltonian can be written in the basis
\begin{equation}
    \Psi_{\text{2-hole}} = 
    \begin{pmatrix}
        b_{\downarrow}^\dagger d_{\uparrow}^\dagger, &
        b_{\uparrow}^\dagger d_{\downarrow}^\dagger, &
        d_{\downarrow}^\dagger d_{\uparrow}^\dagger, &
        b_{\downarrow}^\dagger b_{\uparrow}^\dagger 
    \end{pmatrix}^T |0\rangle.
\end{equation}
In this basis, the Hamiltonian reads
\begin{equation}
    H_{\text{2-hole}} =
    \begin{pmatrix}
        \xi_b & 0 & 2t_{pd} & 2t_{pd} \\
        0 & \xi_b & -2t_{pd} & -2t_{pd} \\
        2t_{pd} & -2t_{pd} & U_d & 0 \\
        2t_{pd} & -2t_{pd} & 0 & 2 \xi_b 
    \end{pmatrix}.
\end{equation}
To separate the singlet and triplet sectors, we introduce the transformed basis
\begin{align}
    |T_{db}\rangle &= \frac{1}{\sqrt{2}} (b_{\downarrow}^\dagger d_{\uparrow}^\dagger + b_{\uparrow}^\dagger d_{\downarrow}^\dagger)|0\rangle, \\
    |S_{db}\rangle &= \frac{1}{\sqrt{2}} (b_{\downarrow}^\dagger d_{\uparrow}^\dagger - b_{\uparrow}^\dagger d_{\downarrow}^\dagger)|0\rangle, \\
    |S_{dd}\rangle &= d_{\downarrow}^\dagger d_{\uparrow}^\dagger |0\rangle, \\
    |S_{bb}\rangle &= b_{\downarrow}^\dagger b_{\uparrow}^\dagger |0\rangle.
\end{align}
In this symmetry-adapted basis $(|T_{db}\rangle, |S_{db}\rangle, |S_{dd}\rangle, |S_{bb}\rangle)^T$, the Hamiltonian becomes block-diagonal:
\begin{equation}
    H_{\text{2-hole}} =
    \begin{pmatrix}
        \xi_b & 0 & 0 & 0 \\
        0 & \xi_b & 2\sqrt{2}t_{pd} & 2\sqrt{2}t_{pd} \\
        0 & 2\sqrt{2}t_{pd} & U_d & 0 \\
        0 & 2\sqrt{2}t_{pd} & 0 & 2 \xi_b 
    \end{pmatrix}.
\end{equation}
The triplet state $|T_{db}\rangle$ completely decouples. For the singlet sector, assuming $U_d \gg \xi_b, t_{pd}$, we can downfold the high-energy Cu doublon state $|S_{dd}\rangle$. The resulting effective $2 \times 2$ Hamiltonian, denoted by $H_{\rm ZR}$, is given in Eq.~\eqref{eq:zr_local_hamiltonian} of the main text for the low-energy singlets $(|S_{db}\rangle, |S_{bb}\rangle)^T$. We denote its lowest eigenvalue by $E_{ZR}$ and the corresponding localized ZR-singlet eigenstate by $|ZR\rangle$. Explicitly,
\begin{equation}
    E_{ZR} = \frac{3\xi_b}{2}-\frac{4 t_{pd}^2}{U_d}
    - \frac{1}{2}\sqrt{\left(\xi_b + \frac{8 t_{pd}^2}{U_d}\right)^2 +32 t_{pd}^2}.
\end{equation}
When the Cu--ligand hybridization dominates the diagonal mismatch, namely $|\xi_b+8t_{pd}^2/U_d|\ll4\sqrt{2}t_{pd}$, the ZR energy can be approximated as
\begin{equation}
    E_{ZR} \approx \frac{3\xi_b}{2} - \frac{4 t_{pd}^2}{U_d} - 2\sqrt{2}t_{pd}.
\end{equation}
In this limit, the ground state reduces to the localized ZR state in Eq.~\eqref{eq:localized_zr_state}, which serves as the reference vacuum for the neighboring Cu spins in the following downfolding.

\subsection{Emergent Frustrating Exchanges $J_2$ and $J_3$ Mediated by the ZR Singlet}

We first consider the third-nearest-neighbor exchange $J_3$ connecting the two Cu sites $d_L$ and $d_R$ on opposite sides of the ZR center, as shown in Fig.~\ref{fig:ZR}(b). The projected Cu--O hybridization is given in Sec.~\ref{sec:j1j2}. For the detailed downfolding below, we define $t_e = t_{pd}/\sqrt{2}$ and $t_a = t_b = t_{pd}/2$.

To transparently demonstrate this mechanism and provide a clear matrix representation, we first project the system onto the dominant $e_1$-orbital subspace. Taking the $|ZR\rangle$ state as the reference vacuum (setting $E_{ZR} \to 0$ for the low-energy projection), the relevant 9-dimensional basis is exactly analogous to the undoped $J_1$ case:
$\Psi_{J_3}^{(e)} = ( 
        d_{R\downarrow}^\dagger d_{L\uparrow}^\dagger, \  
        d_{R\uparrow}^\dagger d_{L\downarrow}^\dagger, \ 
        d_{R\downarrow}^\dagger e_{1\uparrow}^\dagger, \ 
        d_{R\uparrow}^\dagger e_{1\downarrow}^\dagger, \ 
        e_{1\downarrow}^\dagger d_{L\uparrow}^\dagger, \ 
        e_{1\uparrow}^\dagger d_{L\downarrow}^\dagger, \ 
        d_{R\downarrow}^\dagger d_{R\uparrow}^\dagger, \ 
        d_{L\downarrow}^\dagger d_{L\uparrow}^\dagger, \ \\ 
        e_{1\downarrow}^\dagger e_{1\uparrow}^\dagger \ 
        ) |ZR \rangle.$
In this basis, the Hamiltonian takes the block form $$\tilde{H}_e = \begin{pmatrix} 0 & \tilde{T}_{01} & 0 \\ (\tilde{T}_{01})^\dagger & \tilde{H}_{11} & \tilde{T}_{12} \\ 0 & (\tilde{T}_{12})^\dagger & \tilde{H}_{22} \end{pmatrix},$$ 
where $\tilde{T}_{01}=\begin{pmatrix}-t_e&0&t_e&0\\0&-t_e&0&t_e\end{pmatrix}$, 
$\tilde{H}_{11}=\text{diag}(\xi_e,\xi_e,\xi_e,\xi_e)$, 
$\tilde{T}_{12}=\begin{pmatrix}t_e&0&t_e\\-t_e&0&-t_e\\0&-t_e&-t_e\\0&t_e&t_e\end{pmatrix},$ and 
$\tilde{H}_{22}=\begin{pmatrix}U_d&0&0\\0&U_d&0\\0&0&2\xi_e\end{pmatrix}.$
Downfolding this matrix to the lowest energy state up to the fourth order yields the $e_1$-orbital contribution to the superexchange:
\begin{equation}
    J_{3e} = \frac{4t_e^4}{\xi_e^2}\left(\frac{1}{U_d}+\frac{1}{\xi_e}\right) \approx \frac{4t_e^4}{\xi_e^3}.
\end{equation}
Crucially, because the intermediate state $e_{1\downarrow}^\dagger e_{1\uparrow}^\dagger |ZR\rangle$ has a significantly reduced excitation energy ($\sim 2\xi_e$) compared to the Cu doublon ($U_d$), the $1/\xi_e$ term strongly dominates. Representative virtual hopping paths are shown in Fig.~\ref{fig:ZR}(b).

While the $e_1$ channel provides the most intuitive picture and the largest energy scale, a complete calculation must incorporate all virtual pathways through the $a$ and $b$ orbitals, as well as their cross-terms. By exhaustively mapping all charge-transfer pathways up to the fourth order, the total effective exchange $J_3$ can be decomposed into two main contributions: $J_3 = J_U + J_{\xi_p}$. 

\begin{widetext}
The first term, $J_U$, originates from fourth-order processes involving double occupancy on either outer Cu orbital, $d_L$ or $d_R$, and is given by
\begin{align}
J_{U}={}&
\frac{2}{2\xi_b+U_d-E_{ZR}}
\left(
\frac{t_a^2}{2\xi_b+\xi_a-E_{ZR}}
-\frac{t_e^2}{2\xi_b+\xi_e-E_{ZR}}
-\frac{t_b^2}{\xi_b+U_d-E_{ZR}}
\right)^2
\nonumber\\
&+
\frac{2}{\xi_b+U_d-E_{ZR}}
\Bigg[
\left(
\frac{t_a^2}{\xi_b+\xi_a-E_{ZR}}
-\frac{t_e^2}{\xi_b+\xi_e-E_{ZR}}
\right)^2
\nonumber\\
&\qquad+
\left(
\frac{t_a^2}{\xi_b+\xi_a-E_{ZR}}
-\frac{t_e^2}{\xi_b+\xi_e-E_{ZR}}
\right)
\left(
\frac{t_b^2}{2\xi_b-E_{ZR}}
+\frac{t_b^2}{U_d-E_{ZR}}
\right)
-\frac{t_b^4}{(2\xi_b-E_{ZR})(U_d-E_{ZR})}
\Bigg].
\end{align}
Because every such process contains a large energy denominator of order $U_d$, $J_U$ is parametrically suppressed. The dominant second term, $J_{\xi_p}$, arises from processes where holes hop into the ZR center, forming intermediate states with up to four holes in the central plaquette:
\begin{align}
J_{\xi_p}={}&\frac{4t_a^4}{\xi_a^3}+\frac{4t_e^4}{\xi_e^3}
-\frac{4t_a^2t_e^2}{\xi_a\xi_e}
\left(\frac{1}{\xi_a}+\frac{1}{\xi_e}\right)-\frac{t_b^2t_e^2}{\xi_e+2\xi_b-E_{ZR}}
\left(\frac{1}{\xi_e}+\frac{1}{2\xi_b-E_{ZR}}\right)^2+\frac{t_b^2t_a^2}{\xi_a+2\xi_b-E_{ZR}}
\left(\frac{1}{\xi_a}+\frac{1}{2\xi_b-E_{ZR}}\right)^2.
\end{align}
Owing to the localized potential $E_{loc}$, the intermediate states involving the ZR center have significantly reduced excitation energies. As a result, $J_{\xi_p}$ is much larger than $J_U$, making $J_3$ comparable to, or even larger than, the undoped $J_1$.

The derivation naturally extends to the 2nd NN exchange $J_2$ between two Cu spins at a 90-degree angle (e.g., $d_L$ and $d_U$). The fundamental difference lies in the hybridization geometry:
\begin{align}
H_{\text{hyb}}^{(2)}={}&t_{pd}d_L^\dagger
\left(-\frac{1}{\sqrt{2}}e_1+\frac{1}{2}a+\frac{1}{2}b\right) +t_{pd}d_U^\dagger
\left(\frac{1}{\sqrt{2}}e_2+\frac{1}{2}a-\frac{1}{2}b\right)+h.c.
\end{align}
Because $d_L$ and $d_U$ couple to \textit{orthogonal} $e$ orbitals ($e_1$ and $e_2$, respectively), the dominant $e$-orbital channel ($\propto t_e^4/\xi_e^3$) that drives $J_3$ is strictly forbidden for $J_2$. Consequently, $J_2$ relies solely on the $a$ orbital and the heavily suppressed $b$ orbital. This geometric orthogonality naturally explains why $J_2$ is significantly smaller than $J_3$.

The resulting perturbative estimates and their ED benchmarks are summarized in Table~\ref{tab:benchmark} of the main text.
\end{widetext}

\section{Details of the Variational Monte Carlo Method}\label{sec:vmc_appendix}

In this work, the ground state of the effective spin Hamiltonian $H$ [Eq.~\eqref{eq:eff_H}] is investigated using the Variational Monte Carlo (VMC) method. The optimization of the parameterized trial wave function, $|\Psi(\theta)\rangle = \mathcal{P}_G|\psi_0(\theta)\rangle$, is performed using the stochastic reconfiguration (SR) method. Here, $\{\theta\}$ denotes the extensive set of variational parameters embedded in the mean-field state $|\psi_0\rangle$. The Gutzwiller projector $\mathcal{P}_G$ strictly enforces the local single-occupancy constraint, which is essential to project the fermionic mean-field state onto the localized spin-$1/2$ Hilbert space of the effective model.

A key step in the VMC optimization process is computing the variational energy and its gradient with respect to the variational parameters $\{\theta\}$. The variational energy $E(\theta)$ is the expectation value of the Hamiltonian $H$, calculated as:
\begin{equation}
    E(\theta) = \frac{\langle \Psi(\theta) |H| \Psi(\theta) \rangle}{\langle \Psi(\theta) \mid \Psi(\theta) \rangle} = \frac{\sum_C |\Psi(C; \theta)|^2 E_{\mathrm{loc}}(C)}{\sum_C |\Psi(C; \theta)|^2},
\end{equation}
where $|C\rangle$ represents a real-space spin configuration in the $S^z$ basis, and $E_{\mathrm{loc}}(C)$ denotes the local energy for a given configuration:
\begin{equation}
    E_{\mathrm{loc}}(C) = \sum_{C'} \frac{\langle C|H|C' \rangle \Psi(C'; \theta)}{\Psi(C; \theta)}.
\end{equation}
To minimize the energy, we require its gradient, which is given by:
\begin{equation}
    \vec{f} \equiv -\nabla_{\theta} E(\theta) = -2 \Re \left[ \langle (\nabla_{\theta} \ln\Psi)^* E_{\mathrm{loc}} \rangle - \langle E_{\mathrm{loc}} \rangle \langle (\nabla_{\theta} \ln\Psi)^* \rangle \right].
\end{equation}
The expectation values $\langle \dots \rangle$ are computed via standard Markov chain Monte Carlo sampling over the probability distribution $|\Psi(C; \theta)|^2$.

The SR method accelerates convergence by approximating the Hessian of the energy with a positive-definite Hermitian matrix $S$, also known as the quantum geometric tensor. This avoids the high computational cost of computing the true Hessian while providing an improved search direction. The elements of $S$ are given by the covariance matrix of the logarithmic wave function derivatives:
\begin{equation}
    S_{ij} = \langle (\nabla_{\theta_i} \ln\Psi)^* (\nabla_{\theta_j} \ln\Psi) \rangle - \langle \nabla_{\theta_i} \ln\Psi \rangle^* \langle \nabla_{\theta_j} \ln\Psi \rangle.
\end{equation}
With the matrix $S$, the parameter update step $\Delta\vec{\theta}$ at each optimization iteration is determined by solving the following linear system:
\begin{equation}
    \Delta\vec{\theta} = \frac{\Delta\tau}{2} S^{-1} \vec{f},
\end{equation}
where $\Delta\tau$ is a small discrete time step. This iterative procedure is repeated until the variational energy $E(\theta)$ converges. Here, by fixing $J_1=1$, we use $\Delta\tau=0.01$.

\subsection{Variational Wavefunctions and Mean-Field Ansatz}

In our VMC simulations, the trial ground state for the effective spin model is constructed using the Gutzwiller-projected fermionic wavefunction:
\begin{equation}
    |\Psi \rangle = \mathcal{P}_{G} |\phi_{\text{mf}}\rangle,
\end{equation}
where $\mathcal{P}_G = \prod_i (1 - n_{i\uparrow}n_{i\downarrow})$ is the full Gutzwiller projector. In the context of our spin-$1/2$ model, this operator strictly enforces the single-occupancy constraint (i.e., exactly one fermion per intact site), effectively projecting the itinerant fermionic state back onto the localized spin Hilbert space. 

The unprojected state $|\phi_{\text{mf}}\rangle$ is the ground state of a quadratic mean-field Hamiltonian $H_{\text{mf}}$. To accurately capture the complex interplay between the uniform background, the inhomogeneous magnetic order, and the emergent frustrating exchanges around the dopants, we decompose $H_{\text{mf}}$ into three physically distinct parts:
\begin{equation}
    H_{\text{mf}} = H_{\text{bg}} + H_{\text{mag}} + H_{\text{def}}.
\end{equation}

\textbf{1. Uniform Background Terms ($H_{\text{bg}}$):} 
This part describes the global kinetic and pairing background on the intact 1st nearest-neighbor (NN) bonds of the square lattice:
\begin{equation}
    H_{\text{bg}} = \sum_{\langle ij \rangle_1, \sigma} \chi^1 \left(f_{i\sigma}^{\dagger}f_{j\sigma} + h.c.\right) + \mu \sum_{i,\sigma} n_{i\sigma} + \sum_{\langle ij \rangle_1} \left( \Delta_{ij}^1 \hat{B}_{ij}^\dagger + h.c. \right),
\end{equation}
where $\hat{B}_{ij}^\dagger = f_{i\uparrow}^{\dagger}f_{j\downarrow}^{\dagger} - f_{i\downarrow}^{\dagger}f_{j\uparrow}^{\dagger}$ is the singlet pairing operator. The variational parameters here include:
\begin{itemize}
    \item $\chi^1$: The uniform 1st NN hopping amplitude.
    \item $\mu$: The global chemical potential.
    \item $\eta^{s1}$ and $\eta^{d1}$: The extended $s$-wave and $d$-wave pairing amplitudes, respectively. The bond-directional pairing parameter is constructed as $\Delta_{ij}^1 = \eta^{s1} + \eta^{d1}$ for bonds along the $x$-direction, and $\Delta_{ij}^1 = \eta^{s1} - \eta^{d1}$ for bonds along the $y$-direction.
\end{itemize}
It is important to note that, consistent with the effective spin Hamiltonian where the local $J_1$ couplings connected to the dopant sites are broken, the 1st NN hopping and pairing terms are strictly removed (i.e., set to zero) on these specific bonds. The summation $\sum_{\langle ij \rangle_1}$ therefore runs only over the intact bonds of the diluted lattice.

\textbf{2. Inhomogeneous Magnetic Terms ($H_{\text{mag}}$):}
To capture the spatial fragmentation of the AFM order and the emergence of static domain walls induced by the dopants (as shown in Fig.~3 of the main text), we introduce site-dependent staggered magnetization parameters:
\begin{equation}
    H_{\text{mag}} = \frac{1}{2}\sum_{i} m^z_{i} \left(n_{i\uparrow} - n_{i\downarrow}\right).
\end{equation}
\begin{itemize}
    \item $m^z_i$: The local effective Zeeman field strength at site $i$. Unlike uniform mean-field theories, treating $m^z_i$ as an independent variational parameter for every single site is crucial for allowing the system to spontaneously form the inhomogeneous domain-wall structures without any energetic bias.
\end{itemize}

\textbf{3. Defect-Induced Local Terms ($H_{\text{def}}$):}
Because the localized Zhang-Rice singlets introduce emergent 2nd NN ($J_2$) and 3rd NN ($J_3$) superexchanges exclusively around the defect centers, the mean-field ansatz must incorporate corresponding fermionic hopping and pairing channels on these specific local bonds:
\begin{equation}
\begin{split}
H_{\text{def}}={}&\sum_{\langle\langle ll'\rangle\rangle}^{\text{ZR}}
\Big[\chi^{(2)}_{ll'} \left(f_{l\sigma}^{\dagger}f_{l'\sigma}+h.c.\right) 
+\left(\eta^{(2)}_{ll'}\hat{B}_{ll'}^\dagger+h.c.\right)\Big] \nonumber\\
&+\sum_{\langle\langle\langle ll'\rangle\rangle\rangle}^{\text{ZR}}
\Big[\chi^{(3)}_{ll'}
\left(f_{l\sigma}^{\dagger}f_{l'\sigma}+h.c.\right) +\left(\eta^{(3)}_{ll'}\hat{B}_{ll'}^\dagger+h.c.\right)\Big].
\end{split}
\end{equation}
Here, the summations run only over the 2nd NN ($\langle\langle ll'\rangle\rangle$) and 3rd NN ($\langle\langle\langle ll'\rangle\rangle\rangle$) bonds that surround the dopant sites. The variational parameters introduced here are:
\begin{itemize}
    \item $\chi^{(2)}_{ll'}$ and $\chi^{(3)}_{ll'}$: The emergent local hopping amplitudes on the 2nd and 3rd NN bonds around the defects, respectively.
    \item $\eta^{(s2)}_{ll'}$, $\eta^{(d2)}_{ll'}$, $\eta^{(s3)}_{ll'}$, and $\eta^{(d3)}_{ll'}$: The corresponding local $s$-wave and $d$-wave pairing amplitudes. Similar to the background terms, they define the bond-dependent pairings $\eta_{ll'}^{(2)} = \eta^{(s2)}_{ll'} \pm \eta^{(d2)}_{ll'}$ and $\eta_{ll'}^{(3)} = \eta^{(s3)}_{ll'} \pm \eta^{(d3)}_{ll'}$, based on the spatial orientation of the bonds.
\end{itemize}
It is worth noting that, to preserve the local spatial symmetry of the Zhang-Rice center, these parameters are constrained to be identical across all equivalent bonds surrounding the same defect. For instance, while $\chi^{(2)}_{ll'}$ can be optimized as an independent variational parameter for different defects, it takes the exact same value for all four 2nd NN bonds around any given dopant, and so do the parameters on the 3rd NN bonds.

\bibliography{reference}        

@article{Aharony1988,
  title   = {Magnetic phase diagram and magnetic pairing in doped {La}$_2${CuO}$_4$},
  author  = {Aharony, Amnon and Birgeneau, R. J. and Coniglio, A. and Kastner, M. A. and Stanley, H. E.},
  journal = {Phys. Rev. Lett.},
  volume  = {60},
  issue   = {13},
  pages   = {1330--1333},
  year    = {1988},
  doi     = {10.1103/PhysRevLett.60.1330}
}

@article{Alloul2009,
  title     = {Defects in correlated metals and superconductors},
  author    = {Alloul, H. and Bobroff, J. and Gabay, M. and Hirschfeld, P. J.},
  journal   = {Rev. Mod. Phys.},
  volume    = {81},
  issue     = {1},
  pages     = {45--108},
  numpages  = {0},
  year      = {2009},
  month     = {Jan},
  publisher = {American Physical Society},
  doi       = {10.1103/RevModPhys.81.45},
  url       = {https://link.aps.org/doi/10.1103/RevModPhys.81.45}
}

@article{Armitage2010,
  title   = {Progress and perspectives on electron-doped cuprates},
  author  = {Armitage, N. P. and Fournier, P. and Greene, R. L.},
  journal = {Rev. Mod. Phys.},
  volume  = {82},
  issue   = {3},
  pages   = {2421--2487},
  year    = {2010},
  doi     = {10.1103/RevModPhys.82.2421}
}

@article{Bednorz1986,
  author  = {Bednorz, J. G. and M{\"u}ller, K. A.},
  title   = {Possible high {T}$_c$ superconductivity in the {Ba}--{La}--{Cu}--{O} system},
  journal = {Zeitschrift f{\"u}r Physik B Condensed Matter},
  volume  = {64},
  number  = {2},
  pages   = {189--193},
  year    = {1986},
  doi     = {10.1007/BF01303701}
}

@misc{Chen2025arXivA,
  title         = {Half-filled metal and molecular-orbital-mediated pairing in cuprate},
  author        = {Sixuan Chen and Zhiheng Yao and Ning Xia and Shusen Ye and Hongrui Zhang and Jianfa Zhao and Qingqing Liu and Changqing Jin and Shuo Yang and Yayu Wang},
  year          = {2025},
  eprint        = {2509.26472},
  archiveprefix = {arXiv},
  primaryclass  = {cond-mat.supr-con}
}

@article{Emery1987,
  title   = {Theory of high-{${\mathrm{T}}_{\mathrm{c}}$} superconductivity in oxides},
  author  = {Emery, V. J.},
  journal = {Phys. Rev. Lett.},
  volume  = {58},
  issue   = {26},
  pages   = {2794--2797},
  year    = {1987},
  doi     = {10.1103/PhysRevLett.58.2794}
}

@article{Emery1988,
  title   = {Mechanism for high-temperature superconductivity},
  author  = {Emery, V. J. and Reiter, G.},
  journal = {Phys. Rev. B},
  volume  = {38},
  issue   = {7},
  pages   = {4547--4556},
  year    = {1988},
  doi     = {10.1103/PhysRevB.38.4547}
}

@article{EMERY1993597,
  title   = {Frustrated electronic phase separation and high-temperature superconductors},
  journal = {Physica C: Superconductivity},
  volume  = {209},
  number  = {4},
  pages   = {597-621},
  year    = {1993},
  issn    = {0921-4534},
  doi     = {10.1016/0921-4534(93)90581-A},
  url     = {https://www.sciencedirect.com/science/article/pii/092145349390581A},
  author  = {V.J. Emery and S.A. Kivelson}
}

@article{Kane1989,
  title   = {Motion of a single hole in a quantum antiferromagnet},
  author  = {Kane, C. L. and Lee, P. A. and Read, N.},
  journal = {Phys. Rev. B},
  volume  = {39},
  issue   = {10},
  pages   = {6880--6897},
  year    = {1989},
  doi     = {10.1103/PhysRevB.39.6880}
}

@article{Keimer1992,
  title   = {N\'eel transition and sublattice magnetization of pure and doped {La}$_2${CuO}$_4$},
  author  = {Keimer, B. and Aharony, A. and Auerbach, A. and Birgeneau, R. J. and Cassanho, A. and Endoh, Y. and Erwin, R. W. and Kastner, M. A. and Shirane, G.},
  journal = {Phys. Rev. B},
  volume  = {45},
  issue   = {13},
  pages   = {7430--7435},
  year    = {1992},
  doi     = {10.1103/PhysRevB.45.7430}
}

@article{Keimer2015,
  author  = {Keimer, B. and Kivelson, S. A. and Norman, M. R. and Uchida, S. and Zaanen, J.},
  title   = {From quantum matter to high-temperature superconductivity in copper oxides},
  journal = {Nature},
  volume  = {518},
  number  = {7538},
  pages   = {179--186},
  year    = {2015},
  doi     = {10.1038/nature14165}
}

@article{Lee2006,
  title   = {Doping a Mott insulator: Physics of high-temperature superconductivity},
  author  = {Lee, Patrick A. and Nagaosa, Naoto and Wen, Xiao-Gang},
  journal = {Rev. Mod. Phys.},
  volume  = {78},
  issue   = {1},
  pages   = {17--85},
  year    = {2006},
  doi     = {10.1103/RevModPhys.78.17}
}

@misc{Li2024,
  title         = {Bound states in doped charge transfer insulators},
  author        = {Pengfei Li and Yang Shen and Mingpu Qin and Kun Jiang and Jiangping Hu and Tao Xiang},
  year          = {2024},
  eprint        = {2408.00576},
  archiveprefix = {arXiv},
  primaryclass  = {cond-mat.supr-con}
}

@article{Mei_2025,
  author        = {{Mei}, Jiong and {Shi}, Shao-Hang and {Xu}, Ping and {Chen}, Ziyan and {Jin}, Hui-Ke and {Qin}, Mingpu and {Li}, Zi-Xiang and {Jiang}, Kun},
  title         = {{Magnetic electron-hole asymmetry in cuprates: a computational revisit}},
  journal       = {arXiv e-prints},
  year          = 2025,
  month         = nov,
  eid           = {arXiv:2511.15608},
  pages         = {arXiv:2511.15608},
  doi           = {10.48550/arXiv.2511.15608},
  archiveprefix = {arXiv},
  eprint        = {2511.15608},
  primaryclass  = {cond-mat.str-el},
  adsurl        = {https://ui.adsabs.harvard.edu/abs/2025arXiv251115608M}
}

@article{Norman2003,
  author  = {M R Norman and C Pépin},
  title   = {The electronic nature of high temperature cuprate superconductors},
  journal = {Reports on Progress in Physics},
  volume  = {66},
  number  = {10},
  pages   = {1547--1610},
  year    = {2003},
  doi     = {10.1088/0034-4885/66/10/R01}
}

@article{pan2001,
  author        = {Pan, S. H. and O'Neal, J. P. and Badzey, R. L. and Chamon, C. and Ding, H. and Engelbrecht, J. R. and Wang, Z. and Eisaki, H. and Uchida, S. and Gupta, A. K. and Ng, K. -W. and Hudson, E. W. and Lang, K. M. and Davis, J. C.},
  date          = {2001/09/01},
  doi           = {10.1038/35095012},
  id            = {Pan2001},
  isbn          = {1476-4687},
  journal       = {Nature},
  number        = {6853},
  pages         = {282--285},
  title         = {Microscopic electronic inhomogeneity in the high-Tc superconductor {Bi$_2$Sr$_2$CaCu$_2$O$_{8+x}$}},
  url           = {https://doi.org/10.1038/35095012},
  volume        = {413},
  year          = {2001}
}

@article{Sachdev1989,
  title   = {Hole motion in a quantum {N\'eel} state},
  author  = {Sachdev, Subir},
  journal = {Phys. Rev. B},
  volume  = {39},
  issue   = {16},
  pages   = {12232--12247},
  year    = {1989},
  doi     = {10.1103/PhysRevB.39.12232}
}

@article{SchmittRink1988,
  title   = {Spectral Function of Holes in a Quantum Antiferromagnet},
  author  = {Schmitt-Rink, S. and Varma, C. M. and Ruckenstein, A. E.},
  journal = {Phys. Rev. Lett.},
  volume  = {60},
  issue   = {26},
  pages   = {2793--2796},
  year    = {1988},
  doi     = {10.1103/PhysRevLett.60.2793}
}

@article{Sheng1996,
  title     = {Phase String Effect in a Doped Antiferromagnet},
  author    = {Sheng, D. N. and Chen, Y. C. and Weng, Z. Y.},
  journal   = {Phys. Rev. Lett.},
  volume    = {77},
  issue     = {25},
  pages     = {5102--5105},
  numpages  = {0},
  year      = {1996},
  month     = {Dec},
  publisher = {American Physical Society},
  doi       = {10.1103/PhysRevLett.77.5102},
  url       = {https://link.aps.org/doi/10.1103/PhysRevLett.77.5102}
}

@article{Shraiman1988,
  title   = {Mobile Vacancies in a Quantum Heisenberg Antiferromagnet},
  author  = {Shraiman, Boris I. and Siggia, Eric D.},
  journal = {Phys. Rev. Lett.},
  volume  = {61},
  issue   = {4},
  pages   = {467--470},
  year    = {1988},
  doi     = {10.1103/PhysRevLett.61.467}
}

@article{Shraiman1989,
  title     = {Spiral phase of a doped quantum antiferromagnet},
  author    = {Shraiman, Boris I. and Siggia, Eric D.},
  journal   = {Phys. Rev. Lett.},
  volume    = {62},
  issue     = {13},
  pages     = {1564--1567},
  numpages  = {0},
  year      = {1989},
  month     = {Mar},
  publisher = {American Physical Society},
  doi       = {10.1103/PhysRevLett.62.1564},
  url       = {https://link.aps.org/doi/10.1103/PhysRevLett.62.1564}
}

@article{Su1989,
  title   = {Self-consistent renormalized hole motion in a quantum antiferromagnet},
  author  = {Su, Z. B. and Li, Y. M. and Lai, W. Y. and Yu, L.},
  journal = {Phys. Rev. Lett.},
  volume  = {63},
  issue   = {12},
  pages   = {1318--1321},
  year    = {1989},
  doi     = {10.1103/PhysRevLett.63.1318}
}

@article{Wang_2025,
  author        = {{Wang}, Zechao and {Yao}, Fengyu and {Huo}, Yuchen and {Wei}, Zhongxu and {Song}, Zhiyuan and {Ren}, Mingqiang and {Cheng}, Ziyuan and {Jia}, Jinfeng and {Sun}, Yu-Jie and {Xue}, Qi-Kun},
  title         = {{Visualization and manipulation of four-leaf clover-shaped electronic state in cuprate}},
  journal       = {arXiv e-prints},
  year          = 2025,
  month         = jun,
  eid           = {arXiv:2506.21392},
  pages         = {arXiv:2506.21392},
  doi           = {10.48550/arXiv.2506.21392},
  archiveprefix = {arXiv},
  eprint        = {2506.21392},
  primaryclass  = {cond-mat.supr-con},
  adsurl        = {https://ui.adsabs.harvard.edu/abs/2025arXiv250621392W}
}

@article{Wang2002,
  title   = {Inhomogeneous $d$-wave superconducting state of a doped Mott insulator},
  author  = {Wang, Ziqiang and Engelbrecht, Jan R. and Wang, Shancai and Ding, Hong and Pan, Shuheng H.},
  journal = {Phys. Rev. B},
  volume  = {65},
  issue   = {6},
  pages   = {064509},
  year    = {2002},
  doi     = {10.1103/PhysRevB.65.064509}
}

@article{Weng1997,
  title     = {Phase string effect in the $t\ensuremath{-}J$ model: General theory},
  author    = {Weng, Z. Y. and Sheng, D. N. and Chen, Y.-C. and Ting, C. S.},
  journal   = {Phys. Rev. B},
  volume    = {55},
  issue     = {6},
  pages     = {3894--3906},
  numpages  = {0},
  year      = {1997},
  month     = {Feb},
  publisher = {American Physical Society},
  doi       = {10.1103/PhysRevB.55.3894},
  url       = {https://link.aps.org/doi/10.1103/PhysRevB.55.3894}
}

@article{Weng2007,
  title     = {Phase String Theory for Doped Antiferromagnets},
  author    = {Weng, Zheng-Yu},
  journal   = {International Journal of Modern Physics B},
  volume    = {21},
  number    = {05},
  pages     = {773--827},
  year      = {2007},
  publisher = {World Scientific},
  doi       = {10.1142/S0217979207036722},
  url       = {https://doi.org/10.1142/S0217979207036722}
}

@article{xia2025,
  title     = {Unveiling Stripe-Shaped Charge Density Modulations in Doped Mott Insulators},
  author    = {Xia, Ning and Guo, Yuchen and Yang, Shuo},
  journal   = {Phys. Rev. Lett.},
  volume    = {135},
  issue     = {11},
  pages     = {116504},
  numpages  = {7},
  year      = {2025},
  month     = {Sep},
  publisher = {American Physical Society},
  doi       = {10.1103/hmqb-9q91},
  url       = {https://link.aps.org/doi/10.1103/hmqb-9q91}
}

@misc{Ye2023,
  title         = {Visualizing the {Zhang-Rice} singlet, molecular orbitals and pair formation in cuprate},
  author        = {Shusen Ye and Jianfa Zhao and Zhiheng Yao and Sixuan Chen and Zehao Dong and Xintong Li and Luchuan Shi and Qingqing Liu and Changqing Jin and Yayu Wang},
  year          = {2023},
  eprint        = {2309.09260},
  archiveprefix = {arXiv},
  primaryclass  = {cond-mat.supr-con}
}

@article{Zhang1988PRB,
  title   = {Effective Hamiltonian for the superconducting {Cu} oxides},
  author  = {Zhang, F. C. and Rice, T. M.},
  journal = {Phys. Rev. B},
  volume  = {37},
  issue   = {7},
  pages   = {3759--3761},
  year    = {1988},
  doi     = {10.1103/PhysRevB.37.3759}
}

@article{zhao_2025,
  title     = {Composite structure of single-particle spectral function in lightly-doped Mott insulators},
  author    = {Zhao, Jing-Yu and Weng, Zheng-Yu},
  journal   = {Phys. Rev. B},
  volume    = {111},
  issue     = {10},
  pages     = {104502},
  numpages  = {11},
  year      = {2025},
  month     = {Mar},
  publisher = {American Physical Society},
  doi       = {10.1103/PhysRevB.111.104502},
  url       = {https://link.aps.org/doi/10.1103/PhysRevB.111.104502}
}

@article{Chou1993,
  title   = {Magnetic phase diagram of lightly doped ${\mathrm{La}}_{2-x}{\mathrm{Sr}}_{x}{\mathrm{CuO}}_{4}$ from $^{139}$La nuclear quadrupole resonance},
  author  = {Chou, F. C. and Borsa, F. and Cho, J. H. and Johnston, D. C. and Lascialfari, A. and Torgeson, D. R. and Ziolo, J.},
  journal = {Phys. Rev. Lett.},
  volume  = {71},
  issue   = {14},
  pages   = {2323--2326},
  year    = {1993},
  doi     = {10.1103/PhysRevLett.71.2323}
}

@article{Gooding1994,
  title   = {Theory of electron-hole asymmetry in doped {CuO}$_2$ planes},
  author  = {Gooding, R. J. and Vos, K. J. E. and Leung, P. W.},
  journal = {Phys. Rev. B},
  volume  = {50},
  issue   = {17},
  pages   = {12866--12875},
  year    = {1994},
  doi     = {10.1103/PhysRevB.50.12866}
}

@article{Gooding1994_2,
  title     = {Theory of coexisting transverse-spin freezing and long-ranged antiferromagnetic order in lightly doped ${\mathrm{La}}_{2\mathrm{\ensuremath{-}}\mathit{x}}$${\mathrm{Sr}}_{\mathit{x}}$${\mathrm{CuO}}_{4}$},
  author    = {Gooding, R. J. and Salem, N. M. and Mailhot, A.},
  journal   = {Phys. Rev. B},
  volume    = {49},
  issue     = {9},
  pages     = {6067--6073},
  numpages  = {0},
  year      = {1994},
  month     = {Mar},
  publisher = {American Physical Society},
  doi       = {10.1103/PhysRevB.49.6067},
  url       = {https://link.aps.org/doi/10.1103/PhysRevB.49.6067}
}

@article{Gooding1997,
  title     = {Sr impurity effects on the magnetic correlations of ${\mathrm{La}}_{2\mathrm{\ensuremath{-}}\mathrm{x}}$${\mathrm{Sr}}_{\mathrm{x}}$${\mathrm{CuO}}_{4}$},
  author    = {Gooding, R. J. and Salem, N. M. and Birgeneau, R. J. and Chou, F. C.},
  journal   = {Phys. Rev. B},
  volume    = {55},
  issue     = {10},
  pages     = {6360--6371},
  numpages  = {0},
  year      = {1997},
  month     = {Mar},
  publisher = {American Physical Society},
  doi       = {10.1103/PhysRevB.55.6360},
  url       = {https://link.aps.org/doi/10.1103/PhysRevB.55.6360}
}

@article{Matsuda1992,
  title   = {Magnetic order, spin correlations, and superconductivity in single-crystal {Nd}$_{1.85}${Ce}$_{0.15}${CuO}$_{4+\delta}$},
  author  = {Matsuda, M. and Endoh, Y. and Yamada, K. and Kojima, H. and Tanaka, I. and Birgeneau, R. J. and Kastner, M. A. and Shirane, G.},
  journal = {Phys. Rev. B},
  volume  = {45},
  issue   = {21},
  pages   = {12548--12555},
  year    = {1992},
  doi     = {10.1103/PhysRevB.45.12548}
}

@article{Matsuda2000PRB,
  title     = {Static and dynamic spin correlations in the spin-glass phase of slightly doped ${\mathrm{La}}_{2-x}{\mathrm{Sr}}_{x}{\mathrm{CuO}}_{4}$},
  author    = {Matsuda, M. and Fujita, M. and Yamada, K. and Birgeneau, R. J. and Kastner, M. A. and Hiraka, H. and Endoh, Y. and Wakimoto, S. and Shirane, G.},
  journal   = {Phys. Rev. B},
  volume    = {62},
  issue     = {13},
  pages     = {9148--9154},
  numpages  = {7},
  year      = {2000},
  month     = {Oct},
  publisher = {American Physical Society},
  doi       = {10.1103/PhysRevB.62.9148}
}

@article{Fujita2002,
  title = {Static magnetic correlations near the insulating-superconducting phase boundary in ${\mathrm{La}}_{2\ensuremath{-}x}{\mathrm{Sr}}_{x}{\mathrm{CuO}}_{4}$},
  author = {Fujita, M. and Yamada, K. and Hiraka, H. and Gehring, P. M. and Lee, S. H. and Wakimoto, S. and Shirane, G.},
  journal = {Phys. Rev. B},
  volume = {65},
  issue = {6},
  pages = {064505},
  numpages = {7},
  year = {2002},
  month = {Jan},
  publisher = {American Physical Society},
  doi = {10.1103/PhysRevB.65.064505},
  url = {https://link.aps.org/doi/10.1103/PhysRevB.65.064505}
}

@article{Niedermayer1998,
  title     = {Common Phase Diagram for Antiferromagnetism in ${\mathrm{La}}_{2\ensuremath{-}\mathit{x}}{\mathrm{Sr}}_{\mathit{x}}{\mathrm{CuO}}_{4}$ and ${Y}_{1\ensuremath{-}\mathit{x}}{\mathrm{Ca}}_{\mathit{x}}{\mathrm{Ba}}_{2}{\mathrm{Cu}}_{3}{O}_{6}$ as Seen by Muon Spin Rotation},
  author    = {Niedermayer, Ch. and Bernhard, C. and Blasius, T. and Golnik, A. and Moodenbaugh, A. and Budnick, J. I.},
  journal   = {Phys. Rev. Lett.},
  volume    = {80},
  issue     = {17},
  pages     = {3843--3846},
  numpages  = {0},
  year      = {1998},
  month     = {Apr},
  publisher = {American Physical Society},
  doi       = {10.1103/PhysRevLett.80.3843},
  url       = {https://link.aps.org/doi/10.1103/PhysRevLett.80.3843}
}

@article{Tohyama1994,
  title   = {Role of next-nearest-neighbor hopping in the {$t$--$t'$--$J$} model},
  author  = {Tohyama, T. and Maekawa, S.},
  journal = {Phys. Rev. B},
  volume  = {49},
  issue   = {5},
  pages   = {3596--3599},
  year    = {1994},
  doi     = {10.1103/PhysRevB.49.3596}
}

@article{White1992,
  title   = {Density matrix formulation for quantum renormalization groups},
  author  = {White, Steven R.},
  journal = {Phys. Rev. Lett.},
  volume  = {69},
  issue   = {19},
  pages   = {2863--2866},
  year    = {1992},
  doi     = {10.1103/PhysRevLett.69.2863}
}

@article{White1993,
  title   = {Density-matrix algorithms for quantum renormalization groups},
  author  = {White, Steven R.},
  journal = {Phys. Rev. B},
  volume  = {48},
  issue   = {14},
  pages   = {10345--10356},
  year    = {1993},
  doi     = {10.1103/PhysRevB.48.10345}
}

@article{Anderson1959,
  title     = {New Approach to the Theory of Superexchange Interactions},
  author    = {Anderson, P. W.},
  journal   = {Phys. Rev.},
  volume    = {115},
  issue     = {1},
  pages     = {2--13},
  numpages  = {0},
  year      = {1959},
  month     = {Jul},
  publisher = {American Physical Society},
  doi       = {10.1103/PhysRev.115.2},
  url       = {https://link.aps.org/doi/10.1103/PhysRev.115.2}
}

@article{Zaanen1985,
  title   = {Band gaps and electronic structure of transition-metal compounds},
  author  = {Zaanen, J. and Sawatzky, G. A. and Allen, J. W.},
  journal = {Phys. Rev. Lett.},
  volume  = {55},
  issue   = {4},
  pages   = {418--421},
  year    = {1985},
  doi     = {10.1103/PhysRevLett.55.418}
}

@article{Sorella2001,
  title   = {Generalized Lanczos algorithm for variational quantum Monte Carlo},
  author  = {Sorella, Sandro},
  journal = {Phys. Rev. B},
  volume  = {64},
  issue   = {2},
  pages   = {024512},
  year    = {2001},
  doi     = {10.1103/PhysRevB.64.024512}
}

@article{Sorella2007,
  author  = {Sorella, Sandro and Casula, Michele and Rocca, Dario},
  title   = {Weak binding between two aromatic rings: Feeling the van der Waals attraction by quantum Monte Carlo methods},
  journal = {The Journal of Chemical Physics},
  volume  = {127},
  number  = {1},
  pages   = {014105},
  year    = {2007},
  doi     = {10.1063/1.2746035}
}

@article{Luke1990,
  title     = {Magnetic order and electronic phase diagrams of electron-doped copper oxide materials},
  author    = {Luke, G. M. and Le, L. P. and Sternlieb, B. J. and Uemura, Y. J. and Brewer, J. H. and Kadono, R. and Kiefl, R. F. and Kreitzman, S. R. and Riseman, T. M. and Stronach, C. E. and Davis, M. R. and Uchida, S. and Takagi, H. and Tokura, Y. and Hidaka, Y. and Murakami, T. and Gopalakrishnan, J. and Sleight, A. W. and Subramanian, M. A. and Early, E. A. and Markert, J. T. and Maple, M. B. and Seaman, C. L.},
  journal   = {Phys. Rev. B},
  volume    = {42},
  issue     = {13},
  pages     = {7981--7988},
  numpages  = {0},
  year      = {1990},
  month     = {Nov},
  publisher = {American Physical Society},
  doi       = {10.1103/PhysRevB.42.7981},
  url       = {https://link.aps.org/doi/10.1103/PhysRevB.42.7981}
}

@book{StaufferAharony1994,
  author    = {Stauffer, Dietrich and Aharony, Ammon},
  title     = {Introduction to Percolation Theory},
  edition   = {2},
  publisher = {Taylor \& Francis},
  address   = {London},
  year      = {1994},
  doi       = {10.1201/9781315274386}
}

@article{Kastner1998,
  title   = {Magnetic, transport, and optical properties of monolayer copper oxides},
  author  = {Kastner, M. A. and Birgeneau, R. J. and Shirane, G. and Endoh, Y.},
  journal = {Rev. Mod. Phys.},
  volume  = {70},
  issue   = {3},
  pages   = {897--928},
  year    = {1998},
  doi     = {10.1103/RevModPhys.70.897}
}

@article{Binder1981,
  title     = {Finite size scaling analysis of ising model block distribution functions},
  author    = {Binder, K.},
  journal   = {Z. Phys. B},
  volume    = {43},
  pages     = {119--140},
  year      = {1981},
  publisher = {Springer}
}

@article{Binder1986,
  title     = {Spin glasses: Experimental facts, theoretical concepts, and open questions},
  author    = {Binder, K. and Young, A. P.},
  journal   = {Rev. Mod. Phys.},
  volume    = {58},
  issue     = {4},
  pages     = {801--976},
  numpages  = {0},
  year      = {1986},
  month     = {Oct},
  publisher = {American Physical Society},
  doi       = {10.1103/RevModPhys.58.801},
  url       = {https://link.aps.org/doi/10.1103/RevModPhys.58.801}
}

@article{Lee2003,
  title     = {Single Spin and Chiral Glass Transition in Vector Spin Glasses in Three Dimensions},
  author    = {Lee, L. W. and Young, A. P.},
  journal   = {Phys. Rev. Lett.},
  volume    = {90},
  issue     = {22},
  pages     = {227203},
  numpages  = {4},
  year      = {2003},
  month     = {Jun},
  publisher = {American Physical Society},
  doi       = {10.1103/PhysRevLett.90.227203},
  url       = {https://link.aps.org/doi/10.1103/PhysRevLett.90.227203}
}

@article{Julien_2003,
  title   = {Magnetic order and superconductivity in La2-xSrxCuO4: a review},
  journal = {Physica B: Condensed Matter},
  volume  = {329-333},
  pages   = {693-696},
  year    = {2003},
  issn    = {0921-4526},
  doi     = {10.1016/S0921-4526(02)01997-X},
  url     = {https://www.sciencedirect.com/science/article/pii/S092145260201997X},
  author  = {M.-H. Julien}
}

@article{Scalapino1975,
  title   = {Generalized Ginzburg-Landau theory of pseudo-one-dimensional systems},
  author  = {Scalapino, D. J. and Imry, Y. and Pincus, P.},
  journal = {Phys. Rev. B},
  volume  = {11},
  issue   = {5},
  pages   = {2042--2048},
  year    = {1975},
  doi     = {10.1103/PhysRevB.11.2042}
}

@article{Wiesendanger2009,
  title     = {Spin mapping at the nanoscale and atomic scale},
  author    = {Wiesendanger, Roland},
  journal   = {Rev. Mod. Phys.},
  volume    = {81},
  issue     = {4},
  pages     = {1495--1550},
  numpages  = {0},
  year      = {2009},
  month     = {Nov},
  publisher = {American Physical Society},
  doi       = {10.1103/RevModPhys.81.1495},
  url       = {https://link.aps.org/doi/10.1103/RevModPhys.81.1495}
}

@article{Yasuda2005,
  title = {N\'eel Temperature of Quasi-Low-Dimensional Heisenberg Antiferromagnets},
  author = {Yasuda, C. and Todo, S. and Hukushima, K. and Alet, F. and Keller, M. and Troyer, M. and Takayama, H.},
  journal = {Phys. Rev. Lett.},
  volume = {94},
  issue = {21},
  pages = {217201},
  numpages = {4},
  year = {2005},
  month = {Jun},
  publisher = {American Physical Society},
  doi = {10.1103/PhysRevLett.94.217201},
  url = {https://link.aps.org/doi/10.1103/PhysRevLett.94.217201}
}

@article{Sandvik2002,
  title = {Classical percolation transition in the diluted two-dimensional $S=\frac{1}{2}$ Heisenberg antiferromagnet},
  author = {Sandvik, Anders W.},
  journal = {Phys. Rev. B},
  volume = {66},
  issue = {2},
  pages = {024418},
  numpages = {17},
  year = {2002},
  month = {Jul},
  publisher = {American Physical Society},
  doi = {10.1103/PhysRevB.66.024418},
  url = {https://link.aps.org/doi/10.1103/PhysRevB.66.024418}
}

@article{Stoudenmire2012,
  title={Studying two-dimensional systems with the density matrix renormalization group},
  author={Stoudenmire, Edwin M and White, Steven R},
  journal={Annu. Rev. Condens. Matter Phys.},
  volume={3},
  number={1},
  pages={111--128},
  year={2012},
  publisher={Annual Reviews},
  doi={10.1146/annurev-conmatphys-020911-125018},
  url={https://doi.org/10.1146/annurev-conmatphys-020911-125018}
}

@article{Liu2009,
  title = {Impurity-Induced Frustration in Correlated Oxides},
  author = {Liu, Cheng-Wei and Liu, Shiu and Kao, Ying-Jer and Chernyshev, A. L. and Sandvik, Anders W.},
  journal = {Phys. Rev. Lett.},
  volume = {102},
  issue = {16},
  pages = {167201},
  numpages = {4},
  year = {2009},
  month = {Apr},
  publisher = {American Physical Society},
  doi = {10.1103/PhysRevLett.102.167201},
  url = {https://link.aps.org/doi/10.1103/PhysRevLett.102.167201}
}

@article{Liu2013,
  title = {Impurity-induced frustration: Low-energy model of diluted oxides},
  author = {Liu, Shiu and Chernyshev, A. L.},
  journal = {Phys. Rev. B},
  volume = {87},
  issue = {6},
  pages = {064415},
  numpages = {15},
  year = {2013},
  month = {Feb},
  publisher = {American Physical Society},
  doi = {10.1103/PhysRevB.87.064415},
  url = {https://link.aps.org/doi/10.1103/PhysRevB.87.064415}
}

@article{Harter2007,
  title = {Antiferromagnetic correlations and impurity broadening of NMR linewidths in cuprate superconductors},
  author = {Harter, J. W. and Andersen, B. M. and Bobroff, J. and Gabay, M. and Hirschfeld, P. J.},
  journal = {Phys. Rev. B},
  volume = {75},
  issue = {5},
  pages = {054520},
  numpages = {14},
  year = {2007},
  month = {Feb},
  publisher = {American Physical Society},
  doi = {10.1103/PhysRevB.75.054520},
  url = {https://link.aps.org/doi/10.1103/PhysRevB.75.054520}
}

@article{Andersen2007,
  title = {Disorder-Induced Static Antiferromagnetism in Cuprate Superconductors},
  author = {Andersen, Brian M. and Hirschfeld, P. J. and Kampf, Arno P. and Schmid, Markus},
  journal = {Phys. Rev. Lett.},
  volume = {99},
  issue = {14},
  pages = {147002},
  numpages = {4},
  year = {2007},
  month = {Oct},
  publisher = {American Physical Society},
  doi = {10.1103/PhysRevLett.99.147002},
  url = {https://link.aps.org/doi/10.1103/PhysRevLett.99.147002}
}

@article{Schmalian1993,
title = {Doping dependence of local magnetic moments and antiferromagnetism in high-Tc superconductors: Asymmetry between electron and hole doping},
journal = {Solid State Communications},
volume = {86},
number = {2},
pages = {119-122},
year = {1993},
issn = {0038-1098},
doi = {https://doi.org/10.1016/0038-1098(93)90933-E},
url = {https://www.sciencedirect.com/science/article/pii/003810989390933E},
author = {J. Schmalian and G. Baumgärtel and K.-H. Bennemann}
}

@article{Saunders2007,
  title = {Spin Freezing in Geometrically Frustrated Antiferromagnets with Weak Disorder},
  author = {Saunders, T. E. and Chalker, J. T.},
  journal = {Phys. Rev. Lett.},
  volume = {98},
  issue = {15},
  pages = {157201},
  numpages = {4},
  year = {2007},
  month = {Apr},
  publisher = {American Physical Society},
  doi = {10.1103/PhysRevLett.98.157201},
  url = {https://link.aps.org/doi/10.1103/PhysRevLett.98.157201}
}

@article{Andreanov2010,
  title = {Spin-glass transition in geometrically frustrated antiferromagnets with weak disorder},
  author = {Andreanov, A. and Chalker, J. T. and Saunders, T. E. and Sherrington, D.},
  journal = {Phys. Rev. B},
  volume = {81},
  issue = {1},
  pages = {014406},
  numpages = {10},
  year = {2010},
  month = {Jan},
  publisher = {American Physical Society},
  doi = {10.1103/PhysRevB.81.014406},
  url = {https://link.aps.org/doi/10.1103/PhysRevB.81.014406}
}

@article{White2015,
  title = {Doping asymmetry and striping in a three-orbital ${\mathrm{CuO}}_{2}$ Hubbard model},
  author = {White, Steven R. and Scalapino, D. J.},
  journal = {Phys. Rev. B},
  volume = {92},
  issue = {20},
  pages = {205112},
  numpages = {7},
  year = {2015},
  month = {Nov},
  publisher = {American Physical Society},
  doi = {10.1103/PhysRevB.92.205112},
  url = {https://link.aps.org/doi/10.1103/PhysRevB.92.205112}
}

@article{Cui2020,
  title = {Ground-state phase diagram of the three-band Hubbard model from density matrix embedding theory},
  author = {Cui, Zhi-Hao and Sun, Chong and Ray, Ushnish and Zheng, Bo-Xiao and Sun, Qiming and Chan, Garnet Kin-Lic},
  journal = {Phys. Rev. Res.},
  volume = {2},
  issue = {4},
  pages = {043259},
  numpages = {14},
  year = {2020},
  month = {Nov},
  publisher = {American Physical Society},
  doi = {10.1103/PhysRevResearch.2.043259},
  url = {https://link.aps.org/doi/10.1103/PhysRevResearch.2.043259}
}

@article{Hukushima1996,
author = {Hukushima ,Koji and Nemoto ,Koji},
title = {Exchange Monte Carlo Method and  Application to Spin Glass Simulations},
journal = {Journal of the Physical Society of Japan},
volume = {65},
number = {6},
pages = {1604-1608},
year = {1996},
doi = {10.1143/JPSJ.65.1604},
URL = {https://doi.org/10.1143/JPSJ.65.1604},
eprint = {https://doi.org/10.1143/JPSJ.65.1604}
}

@article{Earl2005,
    author = {Earl, David J. and Deem, Michael W.},
    title = {Parallel tempering: Theory, applications, and new perspectives},
    journal = {Physical Chemistry Chemical Physics},
    volume = {7},
    number = {23},
    pages = {3910-3916},
    year = {2005},
    month = {01},
    issn = {1463-9076},
    doi = {10.1039/b509983h},
    url = {https://doi.org/10.1039/b509983h},
    eprint = {https://pubs.rsc.org/cp/article-pdf/7/23/3910/181500/b509983h.pdf},
}

@article{Irkhin1999,
  title = {Self-consistent spin-wave theory of layered Heisenberg magnets},
  author = {Irkhin, V. Yu. and Katanin, A. A. and Katsnelson, M. I.},
  journal = {Phys. Rev. B},
  volume = {60},
  issue = {2},
  pages = {1082--1099},
  numpages = {0},
  year = {1999},
  month = {Jul},
  publisher = {American Physical Society},
  doi = {10.1103/PhysRevB.60.1082},
  url = {https://link.aps.org/doi/10.1103/PhysRevB.60.1082}
}

@article{Kunisada2020,
  title = {Observation of small Fermi pockets protected by clean {$\mathrm{CuO}_2$} sheets of a high-{$T_c$} superconductor},
  author = {Kunisada, So and Isono, Shunsuke and Kohama, Yoshimitsu and Sakai, Shiro and Bareille, C{\'e}dric and Sakuragi, Shunsuke and Noguchi, Ryo and Kurokawa, Kifu and Kuroda, Kenta and Ishida, Yukiaki and Adachi, Shintaro and Sekine, Ryotaro and Kim, Timur K. and Cacho, Cephise and Shin, Shik and Tohyama, Takami and Tokiwa, Kazuyasu and Kondo, Takeshi},
  journal = {Science},
  volume = {369},
  number = {6505},
  pages = {833--838},
  year = {2020},
  doi = {10.1126/science.aay7311}
}

@article{Chen2026,
  title = {Persistent Fermi pockets and robust electron pairing in lightly doped {$\mathrm{CuO}_2$} planes of cuprate superconductors},
  author = {Chen, Hao and Shi, Jumin and Li, Yinghao and Luo, Xiangyu and Chen, Yiwen and Yin, Chaohui and Shu, Yingjie and Zhang, Jiuxiang and Miao, Taimin and Liang, Bo and Zhu, Wenpei and Cai, Neng and Ren, Xiaolin and Lin, Chengtian and Zhang, Shenjin and Wang, Zhimin and Zhang, Fengfeng and Yang, Feng and Peng, Qinjun and Xu, Zuyan and Liu, Guodong and Mao, Hanqing and Li, Xintong and Xiang, Tao and Zhao, Lin and Zhou, X. J.},
  journal = {Nat. Commun.},
  volume = {17},
  pages = {6195},
  year = {2026},
  doi = {10.1038/s41467-026-73765-y}
}

@article{Jeong2026,
  title = {{BCS-BEC} crossover driven by small Fermi pockets of a high-{$T_c$} cuprate superconductor},
  author = {Jeong, Junhyeok and Enomoto, Yamato and Kohama, Yoshimitsu and Nakayama, Tomotaka and Ando, Kotaro and Kurokawa, Kifu and Huh, Soonsang and Yang, Zhuo and Nomura, Toshihiro and Watson, Matthew D. and Kim, Timur K. and Cacho, Cephise and Lin, Chun and Hashimoto, Makoto and Lu, Donghui and Sakai, Shiro and Tohyama, Takami and Tokiwa, Kazuyasu and Kondo, Takeshi},
  journal = {Nat. Commun.},
  volume = {17},
  pages = {4810},
  year = {2026},
  doi = {10.1038/s41467-026-73081-5}
}

@article{Yamada2003,
	title = {Commensurate Spin Dynamics in the Superconducting State of an Electron-Doped Cuprate Superconductor},
	author = {Yamada, K. and Kurahashi, K. and Uefuji, T. and Fujita, M. and Park, S. and Lee, S.-H. and Endoh, Y.},
	journal = {Phys. Rev. Lett.},
	volume = {90},
	issue = {13},
	pages = {137004},
	numpages = {4},
	year = {2003},
	month = {Apr},
	publisher = {American Physical Society},
	doi = {10.1103/PhysRevLett.90.137004},
	url = {https://link.aps.org/doi/10.1103/PhysRevLett.90.137004}
}

@article{Fujita2008,
	title = {Low-Energy Spin Fluctuations in the Ground States of Electron-Doped ${\mathrm{Pr}}_{1\ensuremath{-}x}{\mathrm{LaCe}}_{x}{\mathrm{CuO}}_{4+\ensuremath{\delta}}$ Cuprate Superconductors},
	author = {Fujita, M. and Matsuda, M. and Lee, S.-H. and Nakagawa, M. and Yamada, K.},
	journal = {Phys. Rev. Lett.},
	volume = {101},
	issue = {10},
	pages = {107003},
	numpages = {4},
	year = {2008},
	month = {Sep},
	publisher = {American Physical Society},
	doi = {10.1103/PhysRevLett.101.107003},
	url = {https://link.aps.org/doi/10.1103/PhysRevLett.101.107003}
}

@article{FUJITA2003130,
title = {Neutron-scattering study of spin correlations in electron-doped Pr0.89LaCe0.11CuO4 single crystals},
journal = {Physica C: Superconductivity},
volume = {392-396},
pages = {130-134},
year = {2003},
note = {Proceedings of the 15th International Symposium on Superconductivity (ISS 2002): Advances in Superconductivity XV. Part I},
issn = {0921-4534},
doi = {https://doi.org/10.1016/S0921-4534(03)00750-0},
url = {https://www.sciencedirect.com/science/article/pii/S0921453403007500},
author = {M. Fujita and S. Kuroshima and M. Matsuda and K. Yamada}
}

@article{Wilson2006,
	title = {High-Energy Spin Excitations in the Electron-Doped Superconductor ${\mathrm{Pr}}_{0.88}{\mathrm{LaCe}}_{0.12}{\mathrm{CuO}}_{4\ensuremath{-}\ensuremath{\delta}}$ with ${T}_{c}=21\text{ }\text{ }\mathrm{K}$},
	author = {Wilson, Stephen D. and Li, Shiliang and Woo, Hyungje and Dai, Pengcheng and Mook, H. A. and Frost, C. D. and Komiya, Seiki and Ando, Yoichi},
	journal = {Phys. Rev. Lett.},
	volume = {96},
	issue = {15},
	pages = {157001},
	numpages = {4},
	year = {2006},
	month = {Apr},
	publisher = {American Physical Society},
	doi = {10.1103/PhysRevLett.96.157001},
	url = {https://link.aps.org/doi/10.1103/PhysRevLett.96.157001}
}

@article{Fujita2011,
	author = {Fujita ,Masaki and Shigiya ,Kosuke and Kaminaga ,Junya and Nakagawa ,Masaaki and Enoki ,Masanori and Yamada ,Kazuyoshi},
	title = {Spin Correlations in Electron-Doped High-Tc Superconductor},
	journal = {Journal of the Physical Society of Japan},
	volume = {80},
	number = {Suppl.B},
	pages = {SB029},
	year = {2011},
	doi = {10.1143/JPSJS.80SB.SB029},
	URL = { https://doi.org/10.1143/JPSJS.80SB.SB029},
	eprint = { https://doi.org/10.1143/JPSJS.80SB.SB029}
}

\end{document}